\documentclass[journal,twoside]{IEEEtran}
\usepackage{pifont}
\usepackage{stmaryrd}
\usepackage{amsmath}
\usepackage{amsfonts}
\usepackage{amssymb}
\usepackage{mathrsfs}
\usepackage{tipa}
\usepackage{cases}
\usepackage{bbm}
\usepackage{graphicx}
\usepackage{algorithm}
\usepackage{algorithmic}
\usepackage{caption}
\usepackage{amsfonts}
\usepackage{amssymb}
\usepackage{amsmath}
\usepackage{mathrsfs}
\usepackage{verbatim}
\usepackage{subfigure}
\usepackage{balance}
\usepackage{booktabs}
\usepackage{fancybox}
\usepackage{bm}
\usepackage{extarrows}
\usepackage{algorithm}
\usepackage{algorithmic}
\usepackage{multirow}
\usepackage{graphicx}
\usepackage{epstopdf}
\usepackage{footmisc}
\usepackage{stfloats}
\usepackage{color}
\newtheorem{theorem}{Theorem}
\newtheorem{proposition}{Proposition}
\newtheorem{lemma}{Lemma}
\newtheorem{corollary}{Corollary}
\newtheorem{remark}{Remark}
\begin{document}

\title{Adaptive Full-Duplex Jamming Receiver for Secure D2D Links in Random Networks
\thanks{
The authors are with the
Department of Information and Communications Engineering, and also with the Ministry of Education Key Lab for Intelligent Networks and Network Security, Xi'an Jiaotong University, Xi'an, 710049, Shaanxi, P. R. China. Email: {\tt
xjbswhm@gmail.com, zhaobq@stu.xjtu.edu.cn,zhengtx@mail.xjtu.edu.cn}.
}
\author{
Hui-Ming Wang, \emph{Senior Member, IEEE},  \hspace{0.05in} Bing-Qing Zhao, \hspace{0.05in} and \hspace{0.05in} Tong-Xing Zheng, \emph{Member, IEEE} \hspace{0.02in}
}
}
\maketitle
\begin{abstract}
Device-to-device (D2D) communication raises new transmission secrecy protection challenges, since conventional physical layer security approaches, such as multiple antennas and cooperation techniques, are invalid due to its resource/size constraints. The full-duplex (FD) jamming receiver, which radiates jamming signals to confuse eavesdroppers when receiving the desired signal simultaneously, is a promising candidate. Unlike existing endeavors that assume the FD jamming receiver always improves the secrecy performance compared with the half-duplex (HD) receiver, we show
that this assumption highly depends on the instantaneous residual self-interference cancellation level and may be invalid.
We propose an adaptive jamming receiver operating in a switched FD/HD mode for a D2D link in random networks. Subject to the secrecy outage probability constraint, we optimize the transceiver parameters, such as signal/jamming powers, secrecy rates and mode switch criteria, to maximize the secrecy throughput. Most of the optimization operations are taken off-line and only very limited on-line calculations are required to make the scheme with low complexity. Furthermore, some interesting insights are provided, such as the secrecy throughput is a quasi-concave function. Numerical results are demonstrated to verify our theoretical findings, and to show its superiority compared with the receiver operating in the FD or HD mode only.
\end{abstract}
\begin{IEEEkeywords}
Physical layer security, device-to-device (D2D) communication, stochastic geometry, full-duplex (FD), secrecy outage, secrecy throughput.
\end{IEEEkeywords}

\maketitle

\section{Introduction}

The security of wireless communications has aroused extensive attention in recent years with the boom of mobile communication devices use and the flexibility of wireless networks interconnection. Physical layer security, which exploits the randomness of wireless channels to safeguard wireless communications, has been studied as a complement to conventional cryptography techniques \cite{Wyner1975}-\cite{WangMag2015}. The basic principle for the physical layer security approaches is to ensure that the equivalent channel of the legitimate receiver is ``better'' than that of the eavesdropper, to guarantee a positive secrecy capacity. Following this idea, two promising approaches, i.e.,  multiple-antenna technologies and cooperation/relay technologies, have been widely adopted to achieve this goal, such as multi-antenna beamforming \cite{Mukherjee2011,Shi2015}, artificial noise assisted methods \cite{Liao2011,ANzheng2015}, cooperative beamforming \cite{Jeong2012,WangTVT2015}, cooperative jamming \cite{LDong2010,GZheng2011}, and some hybird schemes \cite{Wang2013,CWang2015}. However, in many scenarios, multiple antennas are unavailable due to the size and complexity constraints of the transmitter, and cooperative schemes ({related to node mobility, synchronization and trustworthiness \cite{WangAsnchro2011}-\cite{wangsocial2018}}) are also overhead-demanding due to their distributed feature. Device-to-device (D2D) communication in Internet of Things (IoT) applications is such a typical scenario  \cite{ArashAsadi2014}-\cite{Haus2017}, where D2D pairs, such as sensors, are usually equipped with a single antenna each and could not cooperation. In these scenarios, the proposed schemes employing multiple antennas or cooperation no longer apply. Thus, protecting information from leakage still remains challenging in the physical layer.

Fortunately, the progress of developing full duplex (FD) radios opens a new window in the aforementioned scenarios \cite{Bharadia2013}-\cite{ZhongshanZhang2015}.  The critical challenge in implementing such an FD node is the presence of self-interference (SI) that leaks from the FD node's output to its input. Owing to the evolution of SI cancelation techniques, SI can be suppressed in the spacial domain \cite{Riihonen2011}, digital circuit domain \cite{Schober2012} and analogy circuit domain \cite{Duarte2012}, respectively{\footnote{With the analog SI cancellation and digital SI cancellation, an FD design would provide a total of $110$ dB SI cancellation at most \cite{Sim2016}. The typical SI would be $-87$ dBm at least for relay systems, small cell systems and D2D communication \cite{Sim2016,3GPPD2D}.}}. The receiver of a D2D pair, a data collector for example, can predominantly improve the security of the communication link by simultaneously sending a jamming signal when receiving the confidential signal from the transmitter. The jamming signal is able to disturb the eavesdropper from wiretapping while it can be eliminated by the SI procedure at the FD receiver itself. In such a way, physical layer secrecy performance is improved. We refer to this idea as the FD jamming receiver in this paper. As the FD transceiver becomes implementation practical, the FD jamming receiver scheme turns to be an alternative physical layer security approach for D2D applications.

The FD jamming receiver has already been reported and discussed from secrecy metrics \cite{Mou2015,GJChen2017}, resource allocations \cite{Abedi2017}-\cite{LChen2017}, and transmission designs \cite{GZheng2013}-\cite{R3Yan2018}, etc. Among these endeavors, only \cite{Mou2015} and \cite{LChen2017} have focused on the single-antenna scenario and could be applied in D2D communications with resource/size constraints. However, the assumption that the instantaneous channel state information (ICSI) of a wiretap channel is perfectly known by legitimate nodes in \cite{LChen2017} is difficult to realize, since the eavesdropper is usually passive. Furthermore, the works mentioned above ignore the scenarios in the presence of multiple malicious eavesdroppers.

When multiple eavesdroppers exist, the secrecy performance strongly relies on the randomly spatial positions of the eavesdroppers and the propagation large-scale path losses. By utilizing the framework of stochastic geometry theory and Poisson point process (PPP) \cite{Haenggi2009}-\cite{ElSawy2017}, studies on physical layer security with the FD jamming receiver against random eavesdroppers have been carried out in \cite{ZhengDWN2017,ZhengHybrid2017}. 
The analysis of \emph{network-wide} secrecy performance, such as the area secure link number and the network secrecy throughput, have been focused on in \cite{ZhengDWN2017} and \cite{ZhengHybrid2017}, where no specific secure transmission scheme has been proposed.

All of the above investigations made a fundamental assumption by default,  that the FD jamming receiver improves the secrecy performance unconditionally compared with the conventional HD receiver. However, we point out that \emph{this conclusion actually highly depends on the efficiency of the SI suppression},  i.e., the value of the instantaneous SI channel gain. We note that the instantaneous SI channel is usually modeled as a Nakagami-m \cite{ZhongshanZhang2015,Alves2013}, Rayleigh \cite{Riihonen2011,JYao2016}, or Ricean \cite{Duarte2012,Everett2014} random variable.
With a high SI channel gain, the residual SI at the receiver is probably larger than jamming signals received at the eavesdroppers after a large scale fading of the wiretap channels. In such a situation, the overall effect of FD jamming is negative to the secrecy performance, and the receiver in
a half-duplex (HD) mode is better than it in an FD mode. It implies that an adaptive FD jamming receiver should be utilized according to the SI cancelation level, i.e., the receiver will adaptively switch between an FD  mode (transmit jamming when receiving signal) and an HD mode (stop transmitting jamming) . It will obviously outperform the existing pure FD jamming strategy.

Furthermore, for a D2D transmitter with limited hardware and power resources, a full adaptive receiver with on-line transmission parameter optimization ability according to all ICSIs is very difficult with on board calculation ability, if not impossible.
Therefore, a low complexity adaptive FD jamming scheme is an interesting and effective approach to improve the secrecy of a D2D link, which motivates this work. To the best of our knowledge, no prior work has considered this.

\subsection{Our Work and Contributions}
In this paper, we propose a low complexity adaptive jamming receiver operating in a switched FD/HD mode according to the instantaneous residual SI channel gain for a D2D link, coexisting with PPP distributed random eavesdroppers.
The novelty and main contributions of this paper can be summarized as follows:

1) For the first time, we propose an adaptive switched FD/HD jamming receiver secure strategy according to the residual SI channel gain. We optimize a threshold as the mode switch criteria, and design transmission schemes for each mode with the low complexity constraints of D2D nodes under consideration.

2)  In both modes, the optimal transceiver parameters, such as the signal power, secrecy rates, and jamming power,  to maximize the secrecy throughput under the secrecy outage probability (SOP) constraint is optimized off-line. Only very limited calculation should be taken on-line to keep computational complexity low. Explicit optimization solutions to the two cases are provided.

3) We provide new insights into the secure transmission design in both modes. Numerical results show that the secrecy throughput of the proposed strategy is superior to those of schemes with a single receiver mode each.

\subsection{Organization and Notations}
The remainder of this paper is organized as follows. In Section \ref{SEC model}, we present the system model and propose an adaptive switched FD/HD jamming receiver secure strategy. In Section \ref{SEC SOP}, we provide the SOP. In Section \ref{SEC Hybrid}, we solve optimization problems of secrecy throughput maximization under the SOP constraint for each mode. In Section \ref{SEC Switch}, we design an adaptive switched FD/HD jamming receiver transmission scheme with off-line and on-line parts. Numerical results are presented in Section \ref{SEC Numberical}, and Section \ref{SEC Conclude} concludes our work.

We use the following notations in this paper: $\mathbb{P}\left\{ \cdot \right\}$, $\mathcal{F}_v\left( \cdot \right)$ and $\mathbb{E}_v\left[ \cdot \right]$ denote probability, the cumulative distribution function (CDF) of $v$ and the mathematical expectation with respect to (w.r.t.) $v$, respectively. $\mathcal{CN}\left(\mu, \sigma^2\right)$ and $\exp\left(\lambda\right)$ denote the circularly symmetric complex Gaussian distribution with mean $\mu$ and variance $\sigma^2$, and the exponential distribution with parameter $\lambda$, respectively. $\left| \cdot \right|$ and $\Gamma\left( \cdot \right)$ denote Euclidean norm and gamma function, respectively. $\log_2\left( \cdot \right)$ and $\ln\left( \cdot \right)$ denote base-2 and natural logarithms, respectively. $\left( \cdot \right)^\ast$,$\left( \cdot \right)^\dag$ and $\left( \cdot \right)^\star$ represent the optimal solutions. $\thicksim$ stands for ``distributed as''.

\section{System Model and Problem Formulation}\label{SEC model}

\subsection{System Model}\label{SEC A}
\begin{figure}
\centering
\subfigure[]{
\label{modelFD} 
\includegraphics[height=4.2cm,width=8cm]{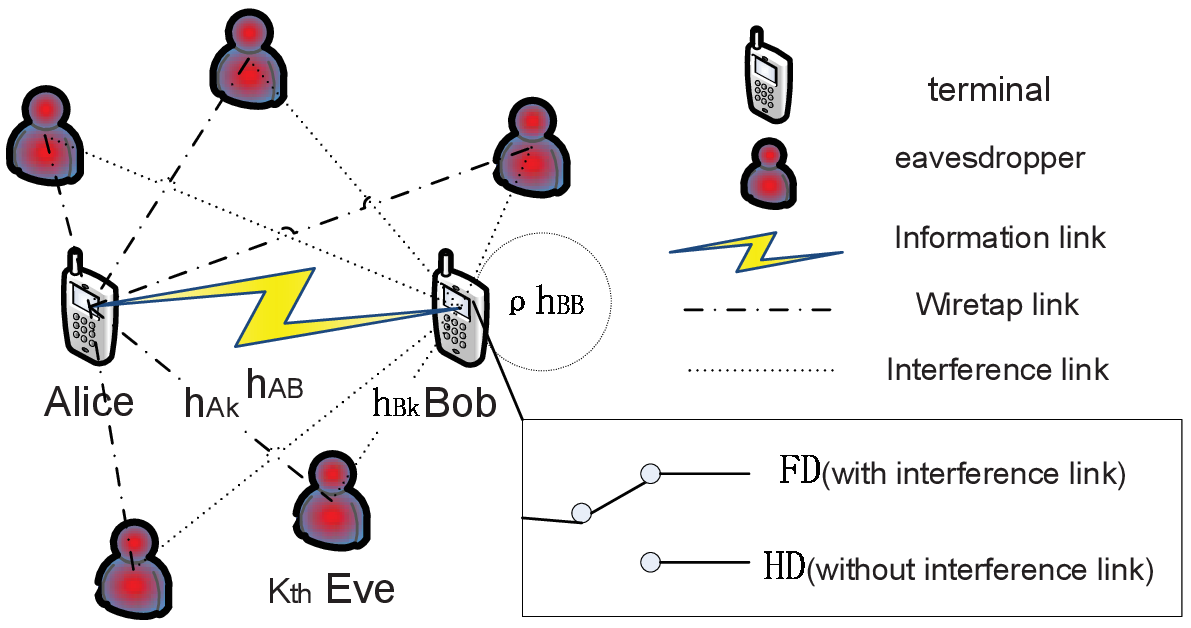}}
\hspace{0.1in}
\subfigure[]{
\label{modelHD} 
\includegraphics[height=4cm,width=4cm]{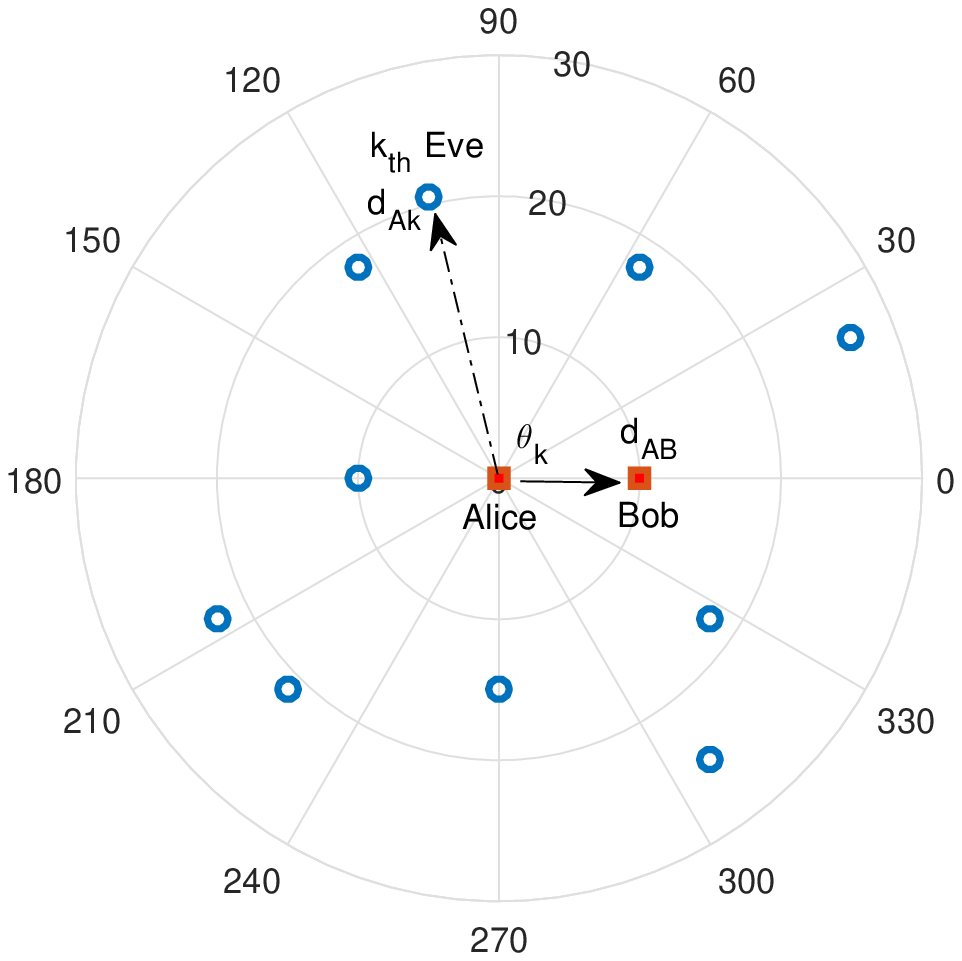}}
\caption{System model with switched FD/HD mode Bob against randomly located Eves.}
\label{figModel} 
\end{figure}

Consider a D2D communication pair depicted in Fig. \ref{figModel}, where a single-antenna transmitter (Alice) delivers confidential information to a single-antenna legitimate receiver (Bob), in the presence of spatially randomly located passive eavesdroppers (Eves). Each of Eves is also equipped with a single antenna. Without loss of generality, we locate Alice at $\left( 0,0 \right)$ and Bob at $\left( d_{AB},0 \right)$ in polar coordinates as shown in Fig. \ref{modelHD}. We model the positions of Eves, $\left\{ e_k:\left( d_{Ak},\theta_k \right) \in \mathbb{R}^{2} \right\}$, as a homogeneous PPP $\Phi$ of intensity $\lambda_e$. The distance between Bob and the $k$-th Eve, $d_{Bk}$, satisfies $d_{Bk}^2 = d_{AB}^2+d_{Ak}^2 - 2d_{AB}d_{Ak}\cos\theta_k$.

For FD Bob, it has the ability to transmit a jamming signal to degrade the quality of the wiretap links while simultaneously receiving the desired signal. The FD mode leads to a feedback loop channel from Bob's output to its input through the channel $\sqrt{\rho}h_{BB}$ shown in Fig. \ref{modelFD}, where $\rho \in \left[ 0,1 \right]$ models the effect of SI suppression in the spacial domain \cite{Sabharwal2014,GZheng2013} and $h_{BB}$ represents the SI channel fading. The value of $\rho$ corresponds to different SI suppression levels, where $\rho=0$ refers to perfect SI suppression and $\rho=1$ means no SI suppression. We will show that the SI suppression level affects the secrecy performance of the FD jamming receiver significantly, i.e., FD mode Bob is not always beneficial to the system security compared with HD mode Bob. To prevent SI from covering the desired signal, Bob switches to operate in the HD mode, i.e., to stop transmit jamming, when the residual SI channel gain is still sufficiently large.

Let $h_{AB}$ denote the channel fading between Alice and Bob. $h_{Ak}$ and $h_{Bk}$ denote the channel fading from Alice and Bob to the $k$-th Eve, respectively. All of the wireless channels undergo quasi-static Rayleigh fading, and $h_{AB},h_{BB},h_{Ak}$ and $h_{Bk}$ are independent and identically distributed (i.i.d.) with zero mean and unit variance \cite{Riihonen2011,Schober2012}, i.e., obeying $\mathcal{CN}\left(0, 1\right)$. The legitimate channel, the wiretap channels and the jamming channels are assumed to suffer from large scale path losses governed by an exponent $\alpha\geq2$. In addition, we assume that the ICSI of Bob is perfectly known by Alice, and the channel state distribution information (CSDI) of Eves is available \footnote{These assumptions are generic in literature on physical layer security, as referred to \cite{ZhengDWN2017,ZhengHybrid2017}.}. The signal received at Bob and the $k$-th Eve is expressed by
\begin{align}
  y_B=\sqrt{P_A}h_{AB}d_{AB}^{-\frac{\alpha}{2}}s_A+\sqrt{P_B}\sqrt{\rho}h_{BB}s_{SI}+n_B, \label{yB}   \\
  y_k=\sqrt{P_A}h_{Ak}d_{Ak}^{-\frac{\alpha}{2}}s_A+\sqrt{P_B}{h_{Bk}}d_{Bk}^{-\frac{\alpha}{2}}s_B+n_k,\label{yk}
\end{align}
where $P_A$ and $P_B$ are the transmit power of the confidential message $s_{A}$ and jamming signal $s_{B}$ with $\mathbb{E}\left[\left|s_{A}\right|^2\right]=1$ and $\mathbb{E}\left[\left|s_{B}\right|^2\right]=1$, respectively. $s_{SI}$ with $\mathbb{E}\left[\left|s_{SI}\right|^2\right]=1$ is the residual SI noise after SI mitigation in the circuit domain \cite{YHua2015,Ahmed2015}.
$n_{B}\thicksim \mathcal{CN}\left(0, \sigma_B^2\right)$ and $n_{k}\thicksim \mathcal{CN}\left(0, \sigma_E^2\right)$ represent the noise at Bob and the $k$-th Eve, respectively.

We consider a wiretap scenario where non-colluding Eves individually decode messages. Hence, the wiretap channel capacity depends on the Eve with the strongest wiretap channel. The capacities of the main channel and of the wiretap channels are then calculated as
\begin{align}
  C_B=\log_2\left(1+\varphi_B \right), \label{CB}   \\
  C_E=\log_2\left(1+\varphi_E \right),\label{CE}
\end{align}
where $\varphi_B \triangleq \frac{P_A\gamma_{AB}d_{AB}^{-\alpha}}{\sigma_B^2+\rho{P_B\gamma_{BB}}}$ and $\varphi_E \triangleq \max\frac{P_A\gamma_{Ak}d_{Ak}^{-\alpha}}{\sigma_E^2+{P_B\gamma_{Bk}d_{Bk}^{-\alpha}}}$ denote the signal-to-interference-plus-noise ratio (SINR) of the main channel and of the equivalent wiretap channel, respectively, with $\gamma_{AB} \triangleq \left|h_{AB}\right|^2$, $\gamma_{BB} \triangleq \left|h_{BB}\right|^2$, $\gamma_{Ak} \triangleq \left|h_{Ak}\right|^2$ and $\gamma_{Bk} \triangleq \left|h_{Bk}\right|^2$. We should note that with $P_B=0$, \eqref{yB}-\eqref{CE} represent the parameters of HD Bob.

\subsection{Secure Transmission Scheme}\label{SEC B}
The well-known Wyner's wiretap encoding scheme is utilized with the codeword rate $R_C$ and the secrecy rate $R_S$. If the main channel can support $R_C$, i.e., $C_B\geq R_C$, Bob can recover the secret message, and the connection is reliable. If the capacity of the wiretap channels, $C_E$, exceeds the redundant rate $R_C-R_S$, perfect secrecy is compromised, and a secrecy outage event occurs.

To avoid an undesired connection outage, i.e., $C_B<R_C$, or an intolerable high secrecy outage, i.e., $C_E\geq{R_C-R_S}$, we propose  the on-off transmission strategy for Alice. Alice transmits only when $\gamma_{AB}$ is not below a preset threshold $\mu_A$, otherwise it keeps silent. Specifically, the transmit power $P_A$ at Alice is adjusted based on $\gamma_{AB}$ and $\gamma_{BB}$ for an FD Bob, and only varies with $\gamma_{AB}$ for HD Bob. We will see that this transmit scheme is with very low complexity.

For FD Bob, the introduction of the jamming signal tends to reduce $C_E$ to enhance the secrecy. When the residual SI channel gain is large, however, the jamming signal leads to more SI than interference to Eves, and Bob is expected to operate in the HD mode. Therefore, we adopt a secure transmission scheme with adaptive switched FD/HD Bob to further safeguard the security. Bob switches between in the FD and HD mode according to the residual SI, i.e., $\rho\gamma_{BB}$. When $\rho\gamma_{BB}$ is sufficiently small, such as $\rho\gamma_{BB}=0$, FD Bob can efficiently interfere with Eves while suffering little SI. With the increasing $\rho\gamma_{BB}$, however, the SI of the FD Bob rises while the jamming to Eves stays unchanged for realization of channels $\gamma_{AB}$, $\gamma_{Ak}$ and $\gamma_{Bk}$. When $\rho\gamma_{BB}$ is large enough, Bob is expected to work in the HD mode to protect the confidential signal against SI. Therefore, a threshold $\mu_B$ of $\rho\gamma_{BB}$ exists, where Bob operates in the FD mode when $\rho\gamma_{BB} < \mu_B$ and in the HD mode when $\rho\gamma_{BB} \geq \mu_B$.

For a given $\mu_B$, we thus have two groups of parameters, i.e., $\left\{ R_C^{FD}, R_S^{FD}, \mu_A^{FD}, P_A^{FD},P_B\right\}$ with FD Bob and $\left\{ R_C^{HD}, R_S^{HD}, \mu_A^{HD}, P_A^{HD} \right\}$ with HD Bob. Owing to the hardware and software constraints of the D2D pair, $R_C^{FD},R_S^{FD},\mu_A^{FD}$ and $R_C^{HD},R_S^{HD},\mu_A^{HD}$ are expected to be optimized off-line and be fixed when transmission takes place, respectively. Bob transmits with an unchanged $P_B$ in the FD mode and with $P_B=0$ in the HD mode. In addition, $P_A^{FD}$ could be adjusted w.r.t. the ICSIs $\gamma_{AB}$ and $\gamma_{BB}$, while $P_A^{HD}$ varies with $\gamma_{AB}$.

\begin{remark}
The proposed scheme adopts a fixed $P_B$ for the FD mode rather than adjustable w.r.t. the ICSIs $\gamma_{AB}$ and $\gamma_{BB}$. Since in the latter case, the accurate optimum solutions to the above two groups of parameters are hard to obtain and they should be solved numerically on-line at any channel realization of $\gamma_{AB}$ and $\gamma_{BB}$, which seems impractical for a D2D communication pair.
\end{remark}

\begin{remark}
$\left\{ R_C^{FD}, R_S^{FD}, \mu_A^{FD}\right\}$, $\left\{ R_C^{HD}, R_S^{HD}, \mu_A^{HD} \right\}$ and $\mu_B$ can be optimized and obtained off-line, which will be shown in the following sections. This greatly reduces the complexity and enhances the operability of our proposed switched FD/HD mode receiver secure strategy.

\end{remark}

\subsection{Performance Metrics}

The SOP \footnote{It describes the secrecy outage performance under specific channel realization, $\gamma_{AB}$, instead of the average performance.} for a given $\gamma_{AB}$ is defined as
\begin{equation}
  \mathcal{P}_{so}\left(\gamma_{AB}\right)\triangleq \mathbb{P}\left\{C_E>R_C - R_S|\gamma_{AB}\right\},\forall\gamma_{AB}\geq\mu_A. \label{SOP}
\end{equation}

To evaluate the average secrecy transmission capacity, we define the secrecy throughput with the fixed $R_S$ as
\begin{equation}
  \Omega_s \triangleq R_S \mathcal{P}_t \mathcal{P}_c, \label{OmegaS0}
\end{equation}
where $\mathcal{P}_t \triangleq \mathbb{P} \left\{\gamma_{AB}\geq\mu_A \right\}$ and $\mathcal{P}_c \triangleq \mathbb{P} \left\{ C_B \geq R_C | \gamma_{AB}\geq\mu_A \right\}$ are the transmission probability of the On-Off scheme, and the conditional connection probability, respectively.

In the following, we optimize the switched FD/HD Bob to maximize the secrecy throughput under the SOP constraint. We start with the analysis of the SOP.

\section{Secrecy Outage Performance Analysis}\label{SEC SOP}

We evaluate the SOP and derive a closed-form approximated expression for the SOP in this section.
{We first give the CDF of the SINR $\varphi_E $ of the equivalent wiretap channel in the following lemma, which would be extensively used in subsequent discussions.}

\begin{lemma} \label{CDF of SINRe}
The CDF of $\varphi_E$ is given by
\begin{align}
\mathcal{F}_{\varphi_E}\left(x\right) &=\exp \Bigg(  - \frac{\lambda_e}{2}\int\nolimits_{0}^{2\pi}\int\nolimits_{0}^{+\infty}
\left(\frac{P_B  }{P_A  } \frac{d_{Ak}^\alpha}{d_{Bk}^\alpha} x+1\right)^{ - 1} \nonumber\\
&\quad \quad \quad \quad \times  \exp\left( -\frac{\sigma_E^2}{P_A }d_{Ak}^\alpha x \right)d d_{Ak}^2 {d\theta_k}  \Bigg) . \label{CDF}
\end{align}
\end{lemma}
\begin{IEEEproof}
Please see Appendix \ref{Appendix.A}.
\end{IEEEproof}

Owing to the double integrals where $d_{Bk}$ is a function of $d_{Ak}$ and $\theta_k$ in \eqref{CDF}, {it is not easy to obtain compact expressions for $\mathcal{F}_{\varphi_E}$ and the SOP $\mathcal{P}_{so}$ related to $\mathcal{F}_{\varphi_E}$}. Note that, to guarantee a reliable communication and to prevent Alice from being overheard, the distance between a D2D pair, $d_{AB}$, is usually small in most applications, {such as $0-0.2$m in Near Field Communications (NFC) systems, $0-10$m in Ultra-wideband (UWB) systems, and $0-30$m in ZigBee and Bluetooth systems \cite{Haus2017,dABFeng2014}}. Therefore, we resort to an asymptotic analysis by considering a small $d_{AB}$ regime as referred to \cite{ZhengDWN2017} and provide a concise approximation for $\mathcal{F}_{\varphi_E}$, which facilitates the analysis of $\mathcal{P}_{so}$ as follows.

\begin{lemma} \label{L CDF}
Let $\beta \triangleq \frac{2\pi}{\alpha}\Gamma\left(\frac{2}{\alpha}\right)$. In the small $d_{AB}$ regime, $\mathcal{F}_{\varphi_E}$ in \eqref{CDF} is approximated by
\begin{equation}
\mathcal{F}_{\varphi_E}\left(x\right) \approx \exp \left( - \beta\lambda_e\left(\frac{P_B}{P_A}x+1\right)^{ - 1}
\left(\frac{\sigma_E^2}{P_A}x \right)^{-\frac{2}{\alpha}} \right). \label{CDFa}
\end{equation}
\end{lemma}
\begin{IEEEproof}
Denoting $\nu\left( x \right)\triangleq d_{Ak}^2 \left(\frac{\sigma_E^2}{P_A}x\right)^\frac{2}{\alpha}$ and substituting $d_{AB}\rightarrow0$ into \eqref{CDF} yield
\begin{align}
&\mathcal{F}_{\varphi_E}\left(x\right) \nonumber\\
& \approx \exp\Bigg( - \pi\lambda_e \left(\frac{P_B}{P_A} x+1\right)^{ - 1}
\left(\frac{\sigma_E^2}{P_A}x \right)^{-\frac{2}{\alpha}}
\nonumber\\
&\quad\quad\quad\quad \times  \int\nolimits_{0}^{+\infty} \exp \left( -\nu^\frac{\alpha}{2} \left( x \right) \right) d\nu\left( x \right) \Bigg) \nonumber\\
&\overset{\left(a\right)}{=}\exp\left( - \frac{2}{\alpha}\Gamma\left( \frac{2}{\alpha} \right)
\pi\lambda_e \left(\frac{P_B}{P_A}x+1\right)^{ - 1}
\left(\frac{\sigma_E^2}{P_A}x \right)^{-\frac{2}{\alpha}}\right),\nonumber
\end{align}
{where $\left(a\right)$ is obtained by calculating the integral w.r.t. $\nu\left( x \right)$ using formula \cite[3.326.1]{mathgraph}}. Replacing $\frac{2\pi}{\alpha}\Gamma\left(\frac{2}{\alpha}\right)$ by $\beta$ completes the proof.
\end{IEEEproof}

Based on results in Lemma \ref{CDF of SINRe} and Lemma \ref{L CDF}, we can obtain both theoretical and approximate expressions for the SOP considering a small regime of $d_{AB}$ in the following theorem.

\begin{theorem} \label{t SOP}
The theoretical expression for the SOP is written as
\begin{align}
&\mathcal{P}_{so}\left(\gamma_{AB}\right)=\nonumber\\
& 1 - \exp\Bigg( { - \frac{\lambda_e}{2}\int\nolimits_{0}^{2\pi}\int\nolimits_{0}^{+\infty}
\left( \left( 2^{R_C-R_S} - 1 \right)\frac{P_B d_{Ak}^\alpha}{P_A\left(\gamma_{AB}\right) d_{Bk}^\alpha}+1\right)^{ - 1} }  \nonumber\\
& \quad \quad \quad \quad \quad  \times  { \exp\left( - \left( 2^{R_C-R_S} - 1 \right)\frac{\sigma_E^2 d_{Ak}^\alpha}{P_A\left(\gamma_{AB}\right)} \right)d d_{Ak}^2{d\theta_k} } \Bigg),\label{SOPf}
\end{align}
and its approximation in a small $d_{AB}$ regime with a closed form is expressed by
\begin{align}
&\mathcal{P}_{so}\left(\gamma_{AB}\right)\approx\nonumber\\
& 1 - \exp\Bigg( - \beta\lambda_e
\left( \left( 2^{R_C-R_S} - 1 \right)\frac{ P_B}{P_A\left(\gamma_{AB}\right)}+1\right)^{ - 1}
\nonumber\\
& \quad \quad \quad \quad \quad  \times \left( \left( 2^{R_C-R_S} - 1 \right)\frac{\sigma_E^2}{P_A\left(\gamma_{AB}\right)} \right)^{-\frac{2}{\alpha}}\Bigg).\label{SOPa}
\end{align}
\end{theorem}

\begin{IEEEproof}
The results can be easily obtained by plugging the CDF $\mathcal{F}_{\varphi_E}$ in \eqref{CDF} and the approximation of the CDF in \eqref{CDFa} into \eqref{SOP}, respectively.
\end{IEEEproof}
We can analyze the relationship between $\mathcal{P}_{so}$ and $\lambda_e$, $R_C$, $R_S$, $P_A$, $P_B$ through either \eqref{SOPf} or \eqref{SOPa}. We find that secrecy outage events less likely occur when enhancing $R_C - R_S$ and $P_B$, or reducing $\lambda_e$ and $P_A$.

\begin{figure}[!t]
\centering
\includegraphics[height=6cm,width=7.5cm ]{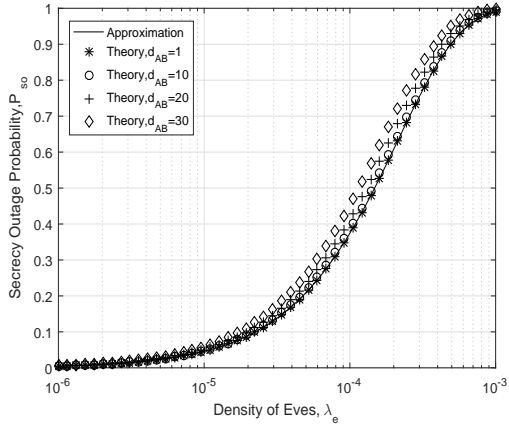}
\caption{{Secrecy outage probability $\mathcal{P}_{so}$ vs. $\lambda_e$ for different values of $d_{AB}$, with $P_A=20$dBm, $P_B=30$dBm, $\sigma^2_E=-90$dBm, $R_C-R_S=3$ bits/s/Hz and $\alpha=4$.}}\label{Fig circa}
\end{figure}

To illustrate the validity of the given approximation method and further to show that the approximation stays accurate even for quite a wide range of $d_{AB}$, we compare the theoretical SOP in \eqref{SOPf} and the approximate SOP in \eqref{SOPa} as shown in Fig. \ref{Fig circa}.
The figure shows that in a wide range of $\lambda_e$ with different orders of magnitude, the theoretical SOP calculated by $d_{AB}$ almost the same as the approximate SOP.
\section{Throughput-Optimal Parameter Design}\label{SEC Hybrid}

In previous sections, we have analyzed the secrecy outage. Next, we aim to maximize the overall secrecy throughput subject to the SOP constraint with switched FD/HD Bob. The overall secrecy throughput is the sum of both modes.
When Bob operates in the FD mode, we adopt a group of fixed $\left\{ R_C^{FD}, R_S^{FD}, \mu_A^{FD}, P_B\right\}$, and adjust $P_A^{FD}$ adaptively w.r.t. the ICSIs of the main channel $\gamma_{AB}$ and the equivalent wiretap channel $\gamma_{BB}$ to maximize the secrecy throughput $\Omega_{FD}$. When Bob works in the HD mode, we adopt another group of invariant $\left\{ R_C^{HD}, R_S^{HD}, \mu_A^{HD} \right\}$ and the optimum expression for $P_A^{HD}$ w.r.t. the ICSI of the main channel $\gamma_{AB}$ to maximize the secrecy throughput $\Omega_{HD}$. Whether Bob operates in the FD or HD mode depends on $\rho\gamma_{BB}$ lower or higher than the operation mode switch threshold $\mu_B$. The overall secrecy throughput is $\Omega_{FD}+\Omega_{HD}$.

We define $\epsilon\in\left(0,1\right)$ as the upper bound of the SOP, and {$\mathcal{P}_{so}^{FD}$ and $\mathcal{P}_{so}^{HD}$ as the approximations of the SOPs for FD and HD mode Bob, respectively. $\mathcal{P}_{so}^{FD}$ is a function w.r.t. $\gamma_{AB}$ and $\gamma_{BB}$ due to $P_A^{FD}\left(\gamma_{AB}, \gamma_{BB}\right)$, while $\mathcal{P}_{so}^{HD}$ is a function w.r.t. $\gamma_{AB}$ due to $P_A^{HD}\left(\gamma_{AB}\right)$}. $P_{Amax}$ and $P_{Bmax}$ are denoted as the maximum transmit power of Alice and Bob, respectively. The optimization problem is formulated as
\begin{subequations}
\label{SubOpt00}
\begin{align}
 &\mathrm{max} \quad
 \Omega_{FD}+\Omega_{HD},\label{OmegaDivi}\\
 &\mathrm{s.t.}\quad \mathcal{P}_{so}^{FD}\left(\gamma_{AB}, \gamma_{BB}\right)\leq\epsilon,\mathcal{P}_{so}^{HD}\left(\gamma_{AB}\right)\leq\epsilon,\label{AllA}\\
 & 0<R_S^{FD}<R_C^{FD}\leq {\log_2\left(1+\frac{P_A^{FD}\left(\gamma_{AB},\gamma_{BB}\right)\gamma_{AB}d_{AB}^{-\alpha}}{\sigma_B^2+\rho P_B\gamma_{BB}} \right)},\nonumber\\ 
 & 0<R_S^{HD}<R_C^{HD}\leq {\log_2\left(1+\frac{P_A^{HD}\left(\gamma_{AB}\right)\gamma_{AB}d_{AB}^{-\alpha}}{\sigma_B^2} \right)},\label{AllC}\\
 & 0<P_A^{FD}\left(\gamma_{AB}, \gamma_{BB}\right)\leq P_{Amax},0<P_A^{HD}\left(\gamma_{AB}\right)\leq P_{Amax},\nonumber\\ 
 & 0<P_B\leq P_{Bmax},\label{AllE}\\
 & \mu_A^{FD}> 0,\mu_A^{HD}> 0,\mu_B\geq 0,\label{AllF}
\end{align}
\end{subequations}
where \eqref{AllA}, \eqref{AllC}, and \eqref{AllE} represent the constraints for secrecy outage, reliable connection, power budgets of Alice and Bob, respectively.

Owing to the reliable connection constraints, i.e., \eqref{AllC}, the secrecy throughput in \eqref{OmegaS0} is transformed into
$\Omega_s= R_S \mathbb{P} \left\{\gamma_{AB}\geq\mu_A \right\}$. Considering the switched FD/HD mode receiver secure strategy, we thus have the concise forms of $\Omega_{FD}$ and $\Omega_{HD}$ as
\begin{align}
  \Omega_{FD}& \triangleq R_S^{FD}\mathbb{P}\left\{\gamma_{AB}\geq\mu_A^{FD},\rho\gamma_{BB}\leq\mu_B\right\} \nonumber\\
& \overset{\left(b\right)}{=}R_S^{FD}\exp\left(-\mu_A^{FD}\right)\left( 1 - \exp\left( - \frac{\mu_B}{\rho}\right) \right),\label{OmegaDi 1}\\
  \Omega_{HD}& \triangleq R_S^{HD}\mathbb{P}\left\{\gamma_{AB}\geq\mu_A^{HD},\rho\gamma_{BB}>\mu_B\right\} \nonumber\\
&\overset{\left(c\right)}{=}R_S^{HD}\exp\left(-\mu_A^{HD}\right)\exp\left( - \frac{\mu_B}{\rho}\right),\label{OmegaDi 2}
\end{align}
where $\left(b\right)$ and $\left(c\right)$ follow from $\gamma_{AB}\thicksim\exp\left(1\right)$ and $\gamma_{BB}\thicksim\exp\left(1\right)$, respectively.

We notice that $\Omega_s$ is a function w.r.t. $\mu_A^{HD}$, $R_S^{HD}$, $R_S^{FD}$, $\mu_A^{FD}$ and $\mu_B$, which are coupled with $R_C^{HD}$, $P_A^{HD}$, $R_C^{FD}$, $P_A^{FD}$ and $P_B$
due to the SOP and reliable connection constraints. Therefore, the objective function \eqref{OmegaDivi} of the optimization problem is provided as
\begin{equation}
 \mathop {\mathrm{max}} \limits_{\substack{\mu_B}} \left(  \mathop {\mathrm{max}} \limits_{\substack{R_C^{FD}, R_S^{FD},P_B\\ \mu_A^{FD}, P_A^{FD}}} \Omega_{FD}+  \mathop {\mathrm{max}} \limits_{\substack{R_C^{HD}, R_S^{HD}\\ \mu_A^{HD},P_A^{HD}}} \Omega_{HD}  \right).\label{stepALL}
\end{equation}

For a given operation mode switch threshold $\mu_B$, the optimization problem in the HD mode can be treated as a special case of that in the FD mode with $P_B=0$. Therefore, we only need to solve the optimization problem in the latter situation, and the optimal solutions of the former situation can be obtained directly. In this section, we perform the optimization procedure step by step for a given $\mu_B$. The optimum $\mu_B$ is solved in the next section.

\emph{For ease of notation, we omit ${FD}$ from $\Omega_{FD},R_C^{FD}, R_S^{FD}$, $\mu_A^{FD}$ $P_A^{FD}$ and $\mathcal{P}_{so}^{FD}$, the optimum solutions or expressions for which are treated as the ones for FD mode Bob by default.} We transform $\Omega$ in \eqref{OmegaDi 1} into $\tilde{\Omega}\triangleq R_S \exp\left( - \mu_A\right)$ for a given $\mu_B$. To maximize $\tilde{\Omega}$, we carry on the equivalent transformation:
\begin{equation}
\mathop {\mathrm{max}} \limits_{\substack{R_C, R_S, \mu_A, P_A, P_B}} \tilde{\Omega} \Longleftrightarrow \mathop {\mathrm{max}} \limits_{\substack{P_B}} \left(   \mathop {\mathrm{max}} \limits_{\substack{R_C, R_S, \mu_A, P_A}} \tilde{\Omega} \right).\label{stepFD}
\end{equation}
Therefore, the entire optimization procedure can be decomposed into two steps: We maximize $\tilde{\Omega}$ by first optimizing over $R_C, R_S$, $\mu_A$ and $P_A$ for a given $P_B$, and then further maximizing the result over the remaining variable $P_B$.

\subsection{Step 1 of the FD Case : Optimum Solutions of $\left\{ R_C, R_S, \mu_A, P_A \right\}$ with a Given $P_B$}
Given a $P_B$, we maximize $\tilde{\Omega}$ over $R_C, R_S$, $\mu_A$ and $P_A$. Due to \eqref{SubOpt00}, a sub-optimization problem is formulated as
\begin{subequations}
\label{SubOptimizationFD}
\begin{align}
 &\mathop {\mathrm{max}} \limits_{\substack{R_C, R_S, \mu_A, P_A}} \quad
 \tilde{\Omega}=R_S\exp\left( - \mu_A\right),\label{FD_A}\\
 &\quad \quad \mathrm{s.t.}\quad \mathcal{P}_{so}\left(\gamma_{AB}, \gamma_{BB}, R_C, R_S\right)\leq\epsilon,\label{FD_B}\\
 &\quad \quad 0<R_S<R_C\leq{\log_2\left(1+\frac{P_A\left(\gamma_{AB},\gamma_{BB}\right)\gamma_{AB}d_{AB}^{-\alpha}}{\sigma_B^2+\rho P_B\gamma_{BB}} \right)},\label{FD_C}\\
 &\quad\quad 0<P_A\left(\gamma_{AB}, \gamma_{BB}\right)\leq{P_{Amax}},\label{FD_D}\\
 &\quad\quad \mu_A\geq{0}.\label{FD_E}
\end{align}
\end{subequations}

Since $\tilde{\Omega}$ in \eqref{FD_A} is a function w.r.t. $R_S$ and $\mu_A$, which are further coupled with $R_C$ and $P_A$ due to the constraints \eqref{FD_B} and \eqref{FD_C}, the problem seems difficult to solve. Treating $\mu_A$ as a function w.r.t. $R_C$ and $R_S$, i.e., $\mu_A\left(R_C, R_S\right)$, we first consider \eqref{FD_C}, \eqref{FD_D} and obtain the expression for $P_A\left(\gamma_{AB},\gamma_{BB} \right)$. Then, we analyze \eqref{FD_B} according to $P_A\left(\gamma_{AB},\gamma_{BB} \right)$. We will show that the constraints in problem \eqref{SubOptimizationFD} will be transformed into an explicit constraint of $\mu_A\left(R_C, R_S\right)$.

Owing to \eqref{FD_C}, we know that to achieve a large $R_S$, $R_C$ is accordingly adjusted to its maximum value, i.e., $R_C={\log_2\left(1+\frac{P_A\left(\gamma_{AB},\gamma_{BB}\right)\gamma_{AB}d_{AB}^{-\alpha}}{\sigma_B^2+\rho P_B\gamma_{BB}} \right)}$. Therefore, we obtain an expression for $P_A$ under $\gamma_{AB}$, $\gamma_{BB}$, $R_C$ and $R_S$:
\begin{align}
&P_A\left(\gamma_{AB},\gamma_{BB}, R_C, R_S \right)=
\nonumber\\
&       \left\{
             \begin{array}{lcl}
             \frac{2^{R_C}-1}{\gamma_{AB}d_{AB}^{-\alpha}}\left( \sigma_B^2+\rho P_B\gamma_{BB} \right), &\gamma_{AB}\geq \mu_A\left( R_C,R_S \right) \\
             0. &\gamma_{AB}<\mu_A\left( R_C,R_S \right)
             \end{array}
        \right.\label{PAfd}
\end{align}
Moreover, with \eqref{PAfd} and \eqref{FD_D}, the feasible range of $P_A\left(\gamma_{AB},\gamma_{BB}, R_C, R_S\right)$ can be expressed by
\begin{align}
&P_A\left(\gamma_{AB},\gamma_{BB}, R_C, R_S\right)\nonumber\\
&\leq \frac{2^{R_C}-1}{\mu_A\left( R_C, R_S\right) d_{AB}^{-\alpha}}\left( \sigma_B^2+ P_B\mu_B \right) \leq P_{Amax}.\label{PAm}
\end{align}
Hence, the transmission is valid only when
\begin{equation}
\mu_A\left( R_C,R_S \right) \geq \mu_{A1} \left( R_C \right) \triangleq \frac{2^{R_C}-1}{P_{Amax}d_{AB}^{-\alpha}}\left( \sigma_B^2+P_B\mu_B \right).\label{mu1FD}
\end{equation}

On the other hand, by plugging $P_A\left(\gamma_{AB},\gamma_{BB}, R_C, R_S\right)$ in \eqref{PAfd} into \eqref{FD_B}, the SOP is
\begin{align}
&\mathcal{P}_{so}\left( \gamma_{AB},\gamma_{BB}, R_C, R_S \right)=\nonumber\\
& 1 - \exp\left( - \beta\lambda_e
\mathcal{G}\left( \gamma_{AB},\gamma_{BB}, R_C, R_S \right)\right),\label{SOPaFD}
\end{align}
where
\begin{align}
&\mathcal{G}\left( \gamma_{AB},\gamma_{BB}, R_C, R_S \right)=\nonumber\\
&\left( \frac{\left( 2^{R_C-R_S}-1 \right) P_B\gamma_{AB}d_{AB}^{-\alpha}}{ \left(2^{R_C}-1\right) \left( \sigma_B^2+\rho P_B\gamma_{BB} \right)}+1\right)^{ - 1} \nonumber\\
& \times  \left( \frac{\left( 2^{R_C-R_S}-1 \right)\sigma_E^2\gamma_{AB}d_{AB}^{-\alpha}}{\left(2^{R_C}-1\right) \left( \sigma_B^2+\rho P_B\gamma_{BB} \right)} \right)^{-\frac{2}{\alpha}}.\label{G}
\end{align}
Obviously, $\mathcal{P}_{so}\left( \gamma_{AB},\gamma_{BB}, R_C, R_S \right)$ monotonically decreases w.r.t. $\gamma_{AB}$ while increasing w.r.t. $\gamma_{BB}$ under given $R_C$ and $R_S$. Since $\gamma_{AB}\geq\mu_A\left( R_C, R_S \right)$ and $\rho\gamma_{BB}\leq\mu_B$, we have
\begin{align}
&\mathop {\mathrm{max}} \limits_{\substack{\gamma_{AB}, \gamma_{BB}}}  \mathcal{P}_{so}\left( \gamma_{AB},\gamma_{BB}, R_C, R_S \right)\nonumber\\
& = 1 - \exp\left( - \beta\lambda_e\mathcal{G}\left( \mu_A\left( R_C, R_S \right),\frac{\mu_B}{\rho}, R_C,R_S \right)\right)\leq\epsilon , \nonumber\\
i.e., &\mathcal{G}\left( \mu_A\left( R_C, R_S \right),\frac{\mu_B}{\rho}, R_C,R_S \right)\leq \tau,\label{SOPm}
\end{align}
where $\tau \triangleq \frac{ - \ln\left( 1-\epsilon \right)}{\beta\lambda_e}$ and $\mathcal{G}\left( \mu_A\left( R_C, R_S \right),\frac{\mu_B}{\rho}, R_C,R_S \right)$ is a monotonic decreasing function w.r.t. $\mu_A\left( R_C, R_S \right)$ under given $R_C$ and $R_S$. Define $\mathcal{G}_1^{-1}\left( x,\frac{\mu_B}{\rho}, R_C,R_S \right)$ as the inverse function of $\mathcal{G}\left( \mu_A\left( R_C, R_S \right),\frac{\mu_B}{\rho}, R_C,R_S \right)$ w.r.t. $\mu_A\left( R_C, R_S \right)$, and we obtain
\begin{equation}
\mu_A\left( R_C,R_S \right) \geq \mu_{A2}\left( R_C,R_S \right) \triangleq \mathcal{G}_1^{-1}\left( \tau, \frac{\mu_B}{\rho}, R_C,R_S \right). \label{mu2FD}
\end{equation}

So far, we have obtained two threshold constraints of $\mu_A$, i.e., \eqref{mu1FD} and \eqref{mu2FD}. Recalling the objective function \eqref{FD_A}, to achieve a large $\tilde{\Omega}$, we expect a small enough $\mu_A\left( R_C,R_S \right)$, which is $\mu_A\left( R_C,R_S \right) = \max\left\{ \mu_{A1}\left( R_C \right),\mu_{A2}\left( R_C,R_S \right)\right\}$ from \eqref{mu1FD} and \eqref{mu2FD}. We discuss the two different values, $\mu_{A1}\left( R_C \right)$ and $\mu_{A2}\left( R_C,R_S \right)$, and obtain a result in the following proposition.

\begin{proposition} \label{p1FD}
The on-off threshold $\mu_{A}\left( R_C,R_S \right)$ satisfies
\begin{equation}
\mu_{A}\left( R_C,R_S \right)=\mu_{A1}\left( R_C \right)= \mu_{A2}\left( R_C,R_S \right). \label{muAFDproposition}
\end{equation}
\end{proposition}
\begin{IEEEproof}
Please see Appendix \ref{Appendix.B}.
\end{IEEEproof}

With this conclusion, the constraints \eqref{FD_B}-\eqref{FD_E} are equivalent to \eqref{muAFDproposition}. From \eqref{muAFDproposition}, we have
\begin{equation}
\left( \frac{\left( 2^{R_C-R_S}-1 \right)P_B }{P_{Amax}}+1\right)^{ - 1}\left( \frac{\left( 2^{R_C-R_S}-1 \right)\sigma_E^2}{P_{Amax}} \right)^{-\frac{2}{\alpha}}= \tau. \label{muOlll}
\end{equation}

By plugging the threshold in \eqref{mu1FD} or \eqref{mu2FD} into \eqref{FD_A} and defining $Y \triangleq 2^{R_C}- 1$, $Z \triangleq \frac{2^{R_C-R_S}-1}{2^{R_C}-1} $, $U \triangleq \frac{ d_{AB}^\alpha }{P_{Amax}}\left( \sigma_B^2+ P_B\mu_B \right)$, the sub-optimization problem \eqref{SubOptimizationFD} is transformed into
\begin{subequations}
\label{SubOptFDsingle}
\begin{align}
 \mathop {\mathrm{max}} \limits_{Y,Z} &\quad
 \tilde{\Omega}= \log_2\left( \frac{1+Y}
 {1+YZ} \right)
 \exp\left( - U Y\right),\label{FD4A}\\
 \mathrm{s.t.} &\left( YZ \frac{P_B }{P_{Amax}}+1\right)^{ - 1}\left( YZ \frac{\sigma_E^2}{P_{Amax}} \right)^{-\frac{2}{\alpha}}= \tau.\label{muOFD}
\end{align}
\end{subequations}
According to \eqref{muOFD}, $Z$ can be treated as a function w.r.t. $Y$, i.e., $Z\left( Y \right)$. Thus, the problem \eqref{SubOptFDsingle} is simplified as a single variable sub-optimization problem:
\begin{equation}
 \mathop {\mathrm{max}} \limits_{Y} \quad \tilde{\Omega}= \log_2\left( \frac{1+Y}
 {1+Y Z\left( Y \right)} \right) \exp\left( - U Y\right).\label{FD41A}
\end{equation}

Although $\tilde{\Omega}$ appears in an implicit function of $Y$, we can still prove that it is quasi-concave on $Y$, and provide the solution to the problem \eqref{FD41A} in the following theorem.
\begin{theorem} \label{t1FD}
Given a $P_B$, the secrecy throughput $\tilde{\Omega}$ in \eqref{FD41A} is a quasi-concave function \cite{CVX2004} of $Y$, and the optimum $Y^{\ast}$ that maximizes $\tilde{\Omega}$ is the unique root of the following equation:
\begin{equation}
\left( V \left( Y \right) \frac{P_B}{P_{Amax}}+1\right)^{ - 1}\left( V \left( Y \right) \frac{\sigma_E^2}{P_{Amax}} \right)^{-\frac{2}{\alpha}}= \tau, \label{t1FDequ}
\end{equation}
where
\begin{equation}
V\left( Y \right) \triangleq \frac{1+Y}{2^{\left( \left(1+Y\right) U \ln2 \right)^{-1}}} - 1.\label{Dfd}
\end{equation}
\end{theorem}
\begin{IEEEproof}
Please see Appendix \ref{Appendix.C}.
\end{IEEEproof}

For a given $P_B$, we can efficiently calculate $Y^\ast$ satisfying \eqref{t1FDequ} by the bisection method, since $\tilde{\Omega}$ is quasi-concave on $Y$. $Z^\ast$ is then obtained by substituting $Y^\ast$ into \eqref{muOFD} and the maximum secrecy throughput $\tilde{\Omega}^\ast$ is also obtained. We thus have the optimum solutions $ R_C^\ast=\log_2\left( 1+Y^\ast \right)$ and $ R_S^\ast=\log_2\left( \frac{1+Y^\ast}{1+Y^\ast Z^\ast} \right)$.
Moreover, we obtain $\mu_A^\ast$ from \eqref{mu1FD} or \eqref{mu2FD} and $P_A^\ast$ from \eqref{PAfd}.

In the following corollary, we further develop some insights into the behavior of $R_C^{\ast}$ and $\tilde{\Omega}^\ast$.
\begin{corollary} \label{c1FD}
The optimum code rate $R_C^{\ast}$ decreases with increasing $\epsilon$ or decreasing $P_{Amax}$ and $\lambda_e$. The maximum secrecy throughput $\tilde{\Omega}^\ast$ increases with increasing $P_{Amax}$ and $\epsilon$ or decreasing $\lambda_e$.
\end{corollary}
\begin{IEEEproof}
Please see Appendix \ref{Appendix.D}.
\end{IEEEproof}
Corollary \ref{c1FD} suggests that to enlarge $\tilde{\Omega}^\ast$, a larger power budget, a looser SOP constraint or a less denser distribution of eavesdroppers should be met.

\subsection{Step 2 of the FD Case : Optimum Solution of $P_B$}
We maximize $\tilde{\Omega}^\ast$ over $P_B$. Owing to \eqref{t1FDequ}, $Y^\ast$ can be treated as a function w.r.t. $P_B$, i.e., $Y^\ast\left( P_B \right)$.
Combining with \eqref{muOFD}, $\log_2\left( \frac{1+Y^\ast\left( P_B \right)}{1+Y^\ast\left( P_B \right)Z\left( Y^\ast\left( P_B \right) \right)} \right)$ in \eqref{FD41A} can be replaced by $\frac{1}{\left(1+Y^\ast\left( P_B \right)\right)U \left( P_B \right)\ln 2}$. The corresponding sub-optimization problem is thus formulated as
\begin{subequations}
\label{SubOptFDPB}
\begin{align}
 \mathop {\mathrm{max}} \limits_{P_B} &\quad
 \tilde{\Omega}^\ast= \frac{\exp\left( - U \left( P_B \right) Y^\ast\left( P_B \right) \right)}{\left(1+Y^\ast\left( P_B \right)\right)U \left( P_B \right)\ln 2}
,\label{FD5A}\\
 \mathrm{s.t.} &\quad 0< P_B \leq P_{Bmax}. \label{FD5G}
\end{align}
\end{subequations}

The following theorem provides the optimum $P_B^\dag$ that maximizes $\tilde{\Omega}^\ast$.
\begin{theorem} \label{t2FD}
The secrecy throughput $\tilde{\Omega}^\ast$ in \eqref{FD5A} is a quasi-concave function of $P_B$, and the optimum $P_B^\dag$ that maximizes $\tilde{\Omega}^\ast$ is
\begin{equation}
P_B^\dag=
       \left\{
             \begin{array}{lcl}
             P_B^\ast, &P_{Bmax}\geq P_B^\ast \\
             P_{Bmax}, &P_{Bmax}<P_B^\ast
             \end{array}
        \right.\label{PBdag}
\end{equation}
where $P_B^\ast$ is the unique root of the following equation:
\begin{align}
&\varpi Y^\ast\left( P_B \right) W \left( Y^\ast\left( P_B \right), P_B \right)\left( 1+V \left( Y^\ast\left( P_B \right), P_B \right) \right) \nonumber\\
&= U \left( P_B \right) V^{2}\left( Y^\ast\left( P_B \right), P_B \right)\left( 1+Y^\ast\left( P_B \right) \right), \label{constraintFD}
\end{align}
and $\varpi \triangleq \frac{ d_{AB}^\alpha }{P_{Amax}}\mu_B$, $W\left( Y^\ast\left( P_B \right), P_B \right) \triangleq \eta P_{Amax}+ \left(1+\eta\right)P_B V \left( Y^\ast\left( P_B \right), P_B \right) $ with $\eta \triangleq \frac{2}{\alpha}$.
\end{theorem}
\begin{IEEEproof}
\emph{For ease of presentation, we omit $P_B$  from $Y^\ast\left( P_B \right)$, $W \left( Y^\ast\left( P_B \right), P_B \right)$, $V\left( Y^\ast\left( P_B \right), P_B \right)$ and $U \left( P_B \right)$, and treat $Y^\ast$, $W\left( Y^\ast \right)$, $V\left( Y^\ast \right)$ and $U$ as functions of $P_B$ by default.}
We first verify that $\tilde{\Omega}^\ast$ is a quasi-concave function of $P_B$ and solve $P_B^\ast$. Then, we compare $P_B^\ast$ with $P_{Bmax}$ to obtain the optimum $P_B^\dag$.

To verify that $\tilde{\Omega}^\ast$ is a quasi-concave function of $P_B$, we are expected to derive that the second-order derivative of $\tilde{\Omega}^\ast$ w.r.t. $P_B$ at the point $P_B^\ast$, where the first-order derivative of $\tilde{\Omega}^\ast$ w.r.t. $P_B$ equals 0, is less than 0.
Before giving the first-/second-order derivative of $\tilde{\Omega}^\ast$ w.r.t. $P_B$, from \eqref{t1FDequ} and \eqref{Dfd}, we calculate $\frac{d Y^\ast}{d P_B}$ using the implicit function derivative rule and $\frac{d V\left( Y^\ast \right)}{d P_B}$ as
\begin{equation}
 \frac{d Y^\ast}{d P_B}
 =-\frac{1+Y^\ast}{U \left( 1+ U\left(1+Y^\ast\right) \right)}\left( \varpi+\frac{U^2 V^2\left( Y^\ast \right) \left(1+Y^\ast\right)}{ W \left( Y^\ast \right) \left( 1+V \left( Y^\ast \right) \right)} \right),\label{dydPB}
\end{equation}
and
\begin{equation}
\frac{d V\left( Y^\ast\right)}{d P_B}
=\left( \left( 1+\frac{1}{U\left( 1+Y^\ast \right)}\right)\frac{d Y^\ast}{d P_B}+\frac{\varpi}{U^2}\right)\frac{ 1+V\left( Y^\ast \right) }{1+ Y^\ast}.\label{dDdPB'}
\end{equation}

Considering \eqref{dydPB}, the first-order derivative of $\tilde{\Omega}^\ast$ w.r.t. $P_B$ from \eqref{FD5A} is provided as
\begin{equation}
\frac{d \tilde{\Omega}^\ast}{d P_B} = \left(\left( \frac{\varpi}{U}+\frac{U V^2\left( Y^\ast \right) \left(1+Y^\ast\right)}{ W \left( Y^\ast \right) \left( 1+V \left( Y^\ast \right) \right)} \right)- \left( Y^\ast+\frac{1}{U} \right) \varpi\right) \tilde{\Omega}^\ast. \label{1dFDOmega}
\end{equation}
We assume there is a variable $P_B^\ast$ satisfying $\frac{d \Omega^\ast}{d P_B}|_{P_B = P_B^\ast}=0$. Thus, we obtain an equality expressed as \eqref{constraintFD}. To verify that $P_B^\ast$ is the unique root of \eqref{constraintFD} and $\tilde{\Omega}^\ast$ achieves its maximum value at $P_B^\ast$, we should consider the sign of second-order derivative of $\tilde{\Omega}^\ast$ w.r.t. $P_B$ at $P_B^\ast$.

Considering \eqref{dydPB}, \eqref{dDdPB'} and \eqref{1dFDOmega}, we have the second-order derivative of $\tilde{\Omega}^\ast$ w.r.t. $P_B$ at $P_B^\ast$ ({$P_B = P_B^\ast$ has been substituted into the following $Y^\ast$, $W\left( Y^\ast \right)$, $V\left( Y^\ast \right)$ and $U$ in this proof.})
\begin{align}
&\frac{d^2 \tilde{\Omega}^\ast}{d {P_B}^2}|_{P_B = P_B^\ast}=-\frac{\tilde{\Omega}^\ast \left(P_B^\ast\right) }{W \left( Y^\ast \right)\left( 1+V\left( Y^\ast\right) \right)}\nonumber\\
&\quad\quad\quad\quad\times \Bigg(\varpi\left(1+\eta\right) Y^\ast \left( 1+V\left( Y^\ast \right) \right)\mathcal{H}_1\left( Y^\ast, P_B^\ast \right)\nonumber\\
&\quad\quad\quad\quad\quad +
\frac{ \varpi V\left( Y^\ast \right)}{Y^\ast\left( 1+\left( 1+Y^\ast \right) U \right)}\mathcal{H}_2\left( Y^\ast, P_B^\ast \right)\Bigg) ,\label{d2OdPB2}
\end{align}
where
\begin{align}
& \mathcal{H}_1\left( Y^\ast, P_B^\ast \right)\triangleq V\left( Y^\ast \right)-\frac{\varpi Y^\ast P_B^\ast}{\left( 1+Y^\ast \right) U} \left( 1+V\left( Y^\ast \right) \right) ,\nonumber\\
& \mathcal{H}_2\left( Y^\ast, P_B^\ast \right)\triangleq 2 Y^{\ast 2}+2 Y^\ast\left( 1+Y^\ast \right)U \left( Y^\ast-V\left( Y^\ast \right) \right) \nonumber\\
&\quad \quad \quad \quad\quad\quad -2\left( 1+Y^\ast\right)V\left( Y^\ast \right)+V\left( Y^\ast\right). \nonumber
\end{align}
Clearly, the sign of $\frac{d^2 \tilde{\Omega}^\ast}{d {P_B}^2}|_{P_B = P_B^\ast}$ is determined by the values of $\mathcal{H}_1\left( Y^\ast, P_B^\ast \right)$ and $\mathcal{H}_2\left( Y^\ast, P_B^\ast\right)$.

The equality in \eqref{constraintFD} can be transformed into $V\left( Y^\ast\right)=\frac{\varpi Y^\ast W \left( Y^\ast \right)}{\left( 1+Y^\ast \right)U V\left( Y^\ast \right)}\left( 1+V \left( Y^\ast \right) \right)$.
With $\eta \in \left( 0,1 \right)$, we have $W\left( Y^\ast \right) \geq P_B^\ast V \left( Y^\ast \right)$ from the expression for $W\left( Y^\ast \right)$. Thus, $\mathcal{H}_1\left( Y^\ast, P_B^\ast \right)$ satisfies
\begin{align}
\mathcal{H}_1\left( Y^\ast, P_B^\ast \right) &\geq \frac{\varpi Y^\ast P_B^\ast V \left( Y^\ast \right)}{\left( 1+Y^\ast \right)U V\left( Y^\ast\right)} \left( 1+V\left( Y^\ast \right) \right) \nonumber\\
&\quad- \frac{\varpi Y^\ast P_B^\ast}{\left( 1+Y^\ast\right) U}\left( 1+V\left( Y^\ast \right) \right)\nonumber\\
&= 0. \label{2dFDc1}
\end{align}
Considering the expression for $V \left( Y^\ast \right)$ in \eqref{Dfd}, we have $Y^\ast > V \left( Y^\ast \right)$ and $\frac{1+Y^\ast}{1 +V\left( Y^\ast \right) }  = 2^{\left( \left(1+Y^\ast\right) U \ln2 \right)^{-1}}$. Due to $\ln x\leq x-1$ with $x\geq1$, the following equivalent transformation holds:
\begin{align}
&\frac{1}{\left( 1+Y^\ast \right)U}\leq\frac{1+Y^\ast}{1+V \left( Y^\ast \right)}-1 \nonumber\\
&\Longleftrightarrow \left( 1+Y^\ast\right)U\left( Y^\ast-V\left( Y^\ast \right) \right)\geq 1+V\left( Y^\ast \right). \label{2dFDc2}
\end{align}
Thus, $\mathcal{H}_2\left( Y^\ast, P_B^\ast \right)$ satisfies
\begin{align}
\mathcal{H}_2\left( Y^\ast, P_B^\ast \right)&\geq 2Y^\ast\left( 1+Y^\ast \right)- V\left( Y^\ast \right) \nonumber\\
&\geq V\left( Y^\ast\right)\left( 2\left( 1+Y^\ast \right)-1 \right)> 0.\label{2dFDc3}
\end{align}

Combining \eqref{d2OdPB2} and \eqref{2dFDc1}, \eqref{2dFDc3}, we have $\frac{d^2 \tilde{\Omega}^\ast}{d {P_B}^2}|_{P_B = P_B^\ast} <0 $. $\tilde{\Omega}^\ast$ in \eqref{FD5A} is a quasi-concave function of $P_B$ and achieves its maximum value at the point $P_B^\ast$.
\begin{algorithm}[htb]
	\caption{Off-line Part of the Switched FD/HD Receiver Scheme}
	\label{scheme1_algorithm}
	\begin{algorithmic}[1]
		\STATE \textbf{Input:} Set $d_{AB}$, $\lambda_e$, $\rho$, $\alpha$, $\sigma_B^2$, $\sigma_E^2$, $\epsilon$, $P_{Amax}$, $P_{Bstep}$, $P_{Bmax}$ and $\mu_B^{step}, \mu_B^{max}$;
        \STATE \textbf{Initialize:} $N_{iter}=1,\Omega_s\left( N_{iter} \right)=0,N_{num}=0$;
        \STATE Solve $R_C^{HD \ast}$ in \eqref{constraintHD} (bisection method), and then we obtain $\Omega_{HD}^\ast$ from \eqref{HDA};
        \FOR{$\mu_B= 0: \mu_B^{step}: \mu_B^{max}$}
            \FOR{$P_B = 0: P_{Bstep}: P_{Bmax}$}
                \STATE Solve $R_C^{FD \ast}$ in \eqref{t1FDequ} (bisection method);
                \IF{$R_C^{FD \ast}$ and $P_B$ satisfy \eqref{constraintFD} }
                    \STATE Break;
                \ENDIF
            \ENDFOR
            \STATE Plug $R_C^{FD \dag}=R_C^{FD \ast}$ and $P_B^{\dag}=P_B$ into \eqref{FD5A} to obtain $\Omega_{FD}^\dag$;
            \STATE $N_{iter}=N_{iter}+1,\Omega_s\left( N_{iter} \right)= \Omega_{FD}^\dag+\Omega_{HD}^\ast$;
            \IF{$\Omega_s\left( N_{iter} \right)\leq \Omega_s\left( N_{iter}-1 \right)$}
                \STATE $\Omega_s\left( N_{iter} \right)=\Omega_s\left( N_{iter}-1 \right)$;
            \ELSE
                \STATE $\mu_B^\star=\mu_B$, $P_B^\star=P_B^\dag$, $R_C^{HD \star }=R_C^{HD \ast }$, $R_C^{FD \star}=R_C^{FD \dag }$;
            \ENDIF
        \ENDFOR
        \STATE $\Omega_s^\star=\Omega_s\left( N_{iter} \right)$;
        \STATE Substitute $\left\{ \mu_B^\star, R_C^{HD \star} \right\}$ and $\left\{ \mu_B^\star, R_C^{FD \star}, P_B^\star \right\}$ into the HD and FD case, respectively.
        \STATE Solve $R_S^{FD \star}$ (bisection method), $R_S^{HD \star}$ from \eqref{muOlll}, $\mu_A^{HD \star}, \mu_A^{FD \star}$ from \eqref{mu1FD}, and $P_A^{HD \star}, P_A^{FD \star}$ from \eqref{PAfd}.
        \STATE \textbf{Output:} The maximum secrecy throughput $\Omega_s^\star$ and the corresponding optimum solutions $\mu_B^\star, P_B^\star$, $R_C^{HD \star}, R_S^{HD \star}, \mu_A^{HD \star}, P_A^{HD \star}, R_C^{FD \star}, R_S^{FD \star}, \mu_A^{FD \star}, P_A^{FD \star}$.
	\end{algorithmic}
\end{algorithm}
Referring to \eqref{FD5G}, we compare $P_B^\ast$ with $P_{Bmax}$. Since $\tilde{\Omega}^\ast$ first increases and then decreases w.r.t. $P_B$, the optimum $P_B^\dag$ is $P_B^\ast$ if $P_{Bmax}\geq P_B^\ast$, or otherwise $P_B^\dag=P_{Bmax}$. The proof is completed.
\end{IEEEproof}

With the above two-step procedure, we have solved the optimization problem for FD Bob. The optimum $P_B^\dag$ is obtained by solving \eqref{constraintFD}, and the maximum $\tilde{\Omega}^\dag$ is achieved by substituting $P_B^\dag$ into $\tilde{\Omega}^\ast$. Since $Y^\ast$ is a function of $P_B$, we know that $R_C^\ast$, $R_S^\ast$, $\mu_A^\ast$ and $P_A^\ast$ are functions of $P_B$. With $P_B^\dag$, we have the optimum solutions $R_C^\dag$, $R_S^\dag$, $\mu_A^\dag$ and $P_A^\dag$.

We develop some insights into $\tilde{\Omega}^\dag$ and $P_B^\dag=P_B^\ast$ in the following corollary.
\begin{corollary}\label{c2FD}
As $\epsilon \rightarrow 1$ or $\lambda_e \rightarrow 0$, $R_C^\ast$ is close to zero for a given $P_B$, and $\tilde{\Omega}^\ast$ decreases w.r.t. $P_B$. Thus, the maximum secrecy throughput $\tilde{\Omega}^\dag$ is obtained at $P_B^\dag=0$.
When $\epsilon \rightarrow 0$ or $\lambda_e \rightarrow \infty$, $P_B^\dag$ decreases with increasing $\epsilon$ or decreasing $\lambda_e$, while $\tilde{\Omega}^\dag$ is totally opposite.
\end{corollary}
\begin{IEEEproof}
Please see Appendix \ref{Appendix.E}.
\end{IEEEproof}
Corollary \ref{c2FD} shows that when the upper bound of the SOP is very small or the eavesdroppers are densely distributed, $\tilde{\Omega}^\dag$ is enhanced. In contrast, $\tilde{\Omega}^\dag$ is obtained at $P_B=0$.

\subsection{HD case}
For HD Bob, considering the secrecy throughput in \eqref{OmegaDi 2}, the optimization problem \eqref{SubOpt00} is transformed into
\begin{subequations}
\label{SubOptimizationHD}
\begin{align}
& \mathop {\mathrm{max}} \limits_{\substack{R_C^{HD}, R_S^{HD}, \mu_A^{HD}, P_A^{HD}}}
 \Omega_{HD}=R_S^{HD}\exp\left( - \mu_A^{HD}\right)\exp\left( - \frac{\mu_B}{\rho}\right) ,\label{HDA}\\
 &\quad \mathrm{s.t.}\quad \mathcal{P}_{so}^{HD}\left(\gamma_{AB},R_C^{HD},R_S^{HD}\right)\leq\epsilon,\\
 &\quad 0<R_S^{HD}<R_C^{HD}\leq{\log_2\left(1+\frac{P_A^{HD}\left(\gamma_{AB}\right)\gamma_{AB}d_{AB}^{-\alpha}}{\sigma_B^2} \right)},\\
 &\quad 0<P_A^{HD}\left(\gamma_{AB}\right)\leq{P_{Amax}},\\
 &\quad \mu_A^{HD}\geq{0}.
\end{align}
\end{subequations}
Through a similar process as Step 1, we obtain the optimum $R_C^{HD \ast}$ as follows.
\begin{corollary} \label{c3FD}
The optimum $R_C^{HD \ast}$ for HD Bob is the unique root of equation:
\begin{equation}
2^{R_C^{HD}}\left( R_C^{HD}-\log_2 \left( 1+\frac{P_{Amax}}{\sigma_E^2}\tau^{-\frac{\alpha}{2}} \right) \right) = \frac{ P_{Amax} }{\sigma_B^2 d_{AB}^\alpha\ln2}.\label{constraintHD}
\end{equation}
\end{corollary}
\begin{IEEEproof}
Substituting $P_B=0$ into \eqref{t1FDequ} of Theorem \ref{t1FD} and replacing $Y$ with $2^{R_C^{HD}}-1$, we obtain the equation \eqref{constraintHD}.
\end{IEEEproof}

We can solve \eqref{constraintHD} to get the optimum $R_C^{HD \ast}$ by the bisection method. $R_S^{HD \ast}$ is then obtained by substituting $R_C=R_C^{HD \ast}$ and $P_B=0$ into \eqref{muOFD} and replacing $R_S$ with $R_S^{HD \ast}$. Moreover, we have the optimum solutions $\mu_A^{HD\ast}$ from \eqref{mu1FD} or \eqref{mu2FD} and $P_A^{HD\ast}$ from \eqref{PAfd}, and the maximum secrecy throughput $\Omega^{HD\ast}$ is also obtained.

\section{Adaptive Switched FD/HD Receiver Scheme }\label{SEC Switch}
So far we have obtained the optimum solutions when Bob works in the FD or HD mode for a given $\mu_B$, and the rest work of \eqref{stepALL} is to calculate the optimum mode switch threshold $\mu_B$. The optimization problem to maximize $\Omega_s$ becomes a single variable optimization problem, which is formulated as
\begin{subequations}
\label{AllOptimization}
\begin{align}
 \mathop {\mathrm{max}} \limits_{\substack{\mu_B}}& \quad
 \Omega_s=\Omega_{FD}^\dag+\Omega_{HD}^\ast,\label{DA}\\
 \mathrm{s.t.} &\quad \mu_B\geq 0.\label{mursopt}
\end{align}
\end{subequations}

Considering that both \eqref{t1FDequ} and \eqref{constraintFD} are implicit equations, the roots $Y^\dag \left( \mu_B \right)$ and $P_B^\dag \left( \mu_B \right)$ are thus implicit and $\Omega_{FD}^\dag= \tilde{\Omega}_{FD}^\dag \left(1-\exp\left(- \frac{\mu_B}{\rho}\right)\right)$ solved from \eqref{FD5A} has no explicit formulation. Moreover, the expression for $\Omega_{HD}^\ast $ resulted from \eqref{HDA} is not explicit due to the implicit root $R_C^{HD \ast} \left( \mu_B \right)$ of \eqref{constraintHD}. The problem to maximize $\Omega_s $ in \eqref{DA} should be solved numerically by a line search over $\mu_B$.

Detailed algorithms to get $\left\{R_C^{HD \star}, R_S^{HD \star}, \mu_A^{HD \star},P_A^{HD \star} \right\}$, $\left\{ R_C^{FD \star}, R_S^{FD \star}, \mu_A^{FD \star}, P_A^{FD \star}, P_B^\star \right\}$ and $\mu_B^\star$ are summarized in Algorithm \ref{scheme1_algorithm}, in which $\mu_B^{step}$ and $\mu_B^{max}$ denote the step size and the maximum value of $\mu_B$ and $P_{Bstep}$ determines the step size of $P_B$. We have to emphasize that all these parameters can be optimized off-line, and are fixed on-line except $P_A^{FD \star}$ and $P_A^{HD \star}$. In Algorithm \ref{scheme2_algorithm}, we provide all the on-line operations of the proposed scheme, which consists of selecting the duplex modes and calculating the powers $P_A^{FD \star}$ or $P_A^{HD \star}$.

\begin{algorithm}[htb]
	\caption{On-line Part of the Switched FD/HD Receiver Scheme}
	\label{scheme2_algorithm}
	\begin{algorithmic}[1]
        \STATE \textbf{Input:} Evaluate $\gamma_{AB}$ and $\gamma_{BB}$;
        \STATE \textbf{Initialize:} The transmitter and the receiver stay silent at the beginning;
            \IF{$\gamma_{BB} \leq \frac{\mu_B^\star}{\rho} $  }
                \IF{$\gamma_{AB} \geq \mu_A^{FD \star} $  }
                    \STATE Substitute $\gamma_{AB}$ and $\gamma_{BB}$ into $P_A^{FD \star}$. The transmitter transmits confidential signal with power $P_A^{FD \star}$, and the receiver transmits jamming signal with power $P_B^\star$;
                    \ENDIF
            \ELSE
                \IF{$\gamma_{AB} \geq \mu_A^{HD \star} $  }
                    \STATE Substitute $\gamma_{AB}$ into $P_A^{HD \star}$. The transmitter transmits confidential signal with power $P_A^{HD \star}$, and the receiver works in the HD mode;
                    \ENDIF
            \ENDIF
    \STATE Loop through On-line Part until the inputs of Off-line Part change or this scheme stops. In the former case, the communication pair restarts executing Off-line Part.
	\end{algorithmic}
\end{algorithm}

\subsection{Complexity Analysis}
The computational consumption of the off-line optimization mainly depends on the processes of the bisection method and the number of loop iterations.
{We first analyze the computational complexity of the three processes of the bisection method.}
Denote $b_i^c$ (resp. $b_i^s$) and $b_p^c$ (resp. $b_p^s$) as the required interval and precision to search $R_C^{HD \ast}$/$R_C^{FD \ast}$ (resp. $R_S^{FD \star}$) with the bisection method, respectively. We need to calculate \eqref{constraintHD} at most $\log_2 \frac{b_i^c}{b_p^c}$ times to search the null point of it. Since the cost of calculating the value of \eqref{constraintHD} is $O\left(2\right)$, the cost of solving $R_C^{HD \ast}$ is $o_1= O\left(2 \log_2 \frac{b_i^c}{b_p^c} \right)$. Similarly, the computational complexities of solving $R_C^{FD \ast}$ from \eqref{t1FDequ} and $R_S^{FD \star}$ from \eqref{muOlll} are $o_2= O\left( 13 \log_2 \frac{b_i^c}{b_p^c} \right)$ and $o_3= O\left( 5 \log_2 \frac{b_i^s}{b_p^s} \right)$, respectively.
Then, we count the number of loop iterations, i.e., $N= \left( \frac{\mu_B^{max}}{\mu_B^{step}} +1 \right) \left( \frac{P_{Bmax}}{P_{Bstep}} +1 \right)$.
Therefore, the computational complexity of the off-line optimization equals about $o_1+No_2+o_3$, i.e., $ O\left(2 \log_2 \frac{b_i^c}{b_p^c} + 13 \left( \frac{\mu_B^{max}}{\mu_B^{step}} +1 \right) \left( \frac{P_{Bmax}}{P_{Bstep}} +1 \right) \log_2 \frac{b_i^c}{b_p^c} + 5 \log_2 \frac{b_i^s}{b_p^s} \right)$.

The computational consumption of the on-line operation with specific channel realizations equals about $O\left( 1 \right)$, which is ignorable compared with the one of the off-line optimization.

\subsection{Metrics for Comparison}

To verify that the secrecy transmission scheme with a switched FD/HD receiver is superior to which only using an HD or FD receiver, we consider the following metrics.

We denote the secrecy throughputs for FD and HD modes when $\rho\gamma_{BB}\leq \mu_B$ as
\begin{align}
&\Omega_{FD}^{comp} \triangleq R_S^{FD }\exp\left( - \mu_A^{FD }\right) \left(1-\exp\left(-\frac{\mu_B}{\rho}\right)\right),\label{OFDcomp}\\
&\Omega_{HD}^{comp} \triangleq R_S^{HD }\exp\left( - \mu_A^{HD }\right)\left(1-\exp\left(-\frac{\mu_B}{\rho}\right)\right),\label{OHDcomp}
\end{align}
for fair comparisons. We further compare the probability of Bob operating in the FD case and in the HD case for our proposed scheme, which are defined as
\begin{align}
\mathcal{P}_{FD} & \triangleq \exp\left(-\mu_A^{FD}\right)\left( 1 - \exp\left( - \frac{\mu_B}{\rho}\right) \right),\label{ProFD}\\
\mathcal{P}_{HD} & \triangleq \exp\left(-\mu_A^{HD}\right)\exp\left( - \frac{\mu_B}{\rho}\right).\label{ProHD}
\end{align}
In the next section, $\left\{\Omega_{FD}^{comp}, \Omega_{HD}^{comp}\right\}$ and $\left\{\mathcal{P}_{FD}, \mathcal{P}_{HD}\right\}$ are analyzed numerically.

\section{Numerical Results}\label{SEC Numberical}
In this section, we present several numerical examples to validate our theoretical analysis. In all simulation experiments, we preset the path loss exponent $\alpha=4$, the distance between Alice and Bob $d_{AB}=10 m$, and the noise variances at Bob and Eves $\sigma_B^2=\sigma_E^2=-90dBm$.

\subsection{Secrecy Throughput Optimization with HD Receiver }
\begin{figure}[!t]
\begin{center}
\centering
\includegraphics[height=5.5cm,width=7cm]{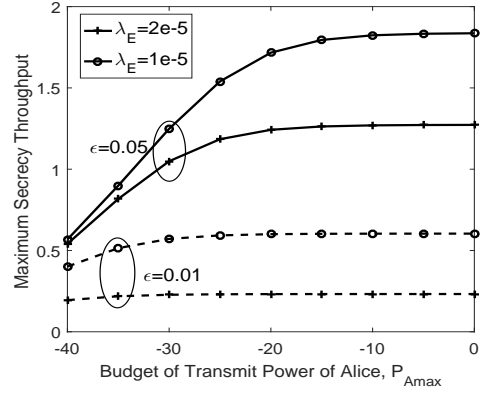}
\end{center}
\caption{Maximum secrecy throughput $\Omega_{HD}^\ast$ vs. $P_{Amax}$ (dBm) for different $\lambda_e's$ and $\epsilon's$, with $\rho=\mu_B=-70$dBm.}\label{Fig1HD}
\end{figure}

Fig. \ref{Fig1HD} depicts the maximum secrecy throughput $\Omega_{HD}^\ast$ solved from \eqref{HDA} versus the transmit power budget of Alice $P_{Amax}$ for different values of $\lambda_e$ and $\epsilon$ when Bob operates in the HD mode for a given $\mu_B$. The relationship between $\Omega_{HD}^\ast$ and $P_{Amax}$, $\lambda_e$, $\epsilon$ in Corollary \ref{c1FD} is validated here. $\Omega_{HD}^\ast$ increases with the growth of small $P_{Amax}$, for a reliable link is more likely to developed between Alice and Bob. With increasing $P_{Amax}$, a secrecy outage event is more likely to occur. When $P_{Amax}$ is sufficiently large, $\Omega_{HD}^\ast$ reaches a plateau since there is a balance between the reliable connection probability and the secrecy outage probability.

\subsection{Secrecy Throughput Optimization with FD Receiver }
\begin{figure}
\centering
\subfigure[]{
\label{Fig1FD} 
\includegraphics[height=5.5cm,width=7.1cm]{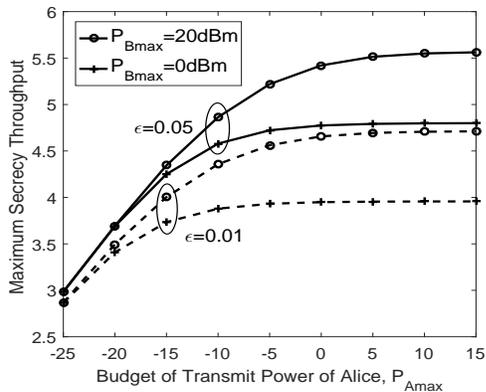}}
\hspace{0.1in}
\subfigure[]{
\label{Fig2FD} 
\includegraphics[height=5.2cm,width=6.8cm]{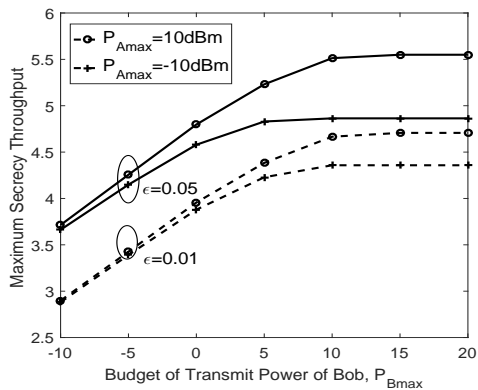}}
\caption{Maximum secrecy throughput $\Omega_{FD}^\dag$  vs. $P_{Amax}$ (dBm) with $P_{Bmax}=0$ or $20$dBm (left) and $P_{Bmax}$ (dBm) with $P_{Amax}= -10$ or $10$ dBm (right), for different values of $\epsilon$ with $\lambda_e=10^{-4}$ and $\rho=\mu_B=-70$dBm.}
\label{figFD} 
\end{figure}

Fig. \ref{figFD} illustrates how the maximum secrecy throughput $\Omega_{FD}^\dag$ solved from \eqref{FD5A} varies w.r.t. the transmit power budgets $P_{Amax}$ and $P_{Bmax}$ for different $\epsilon$. $\Omega_{FD}^\dag$ increases w.r.t. $\epsilon$ as Corollary \ref{c2FD} shows. Similarly to the performance of $\Omega_{HD}^\ast$, $\Omega_{FD}^\dag$ first increases and then almost stays static with increasing $P_{Amax}$ as shown in Fig. \ref{Fig1FD}. Fig. \ref{Fig2FD} shows that the maximum secrecy throughput first increases w.r.t. $P_{Bmax}$ since the jamming signal confuses Eves effectively. When SI is gradually dominant, a certain $P_{Bmax}$ exists to balance the impact of SI and jamming to Eves and $\Omega_{FD}^\dag$ stays unchanged.

\subsection{Maximum Secrecy Throughput with Switched FD/HD Receiver }
\begin{figure}[!t]
\begin{center}
\centering
\includegraphics[height=5.3cm,width=6.6cm]{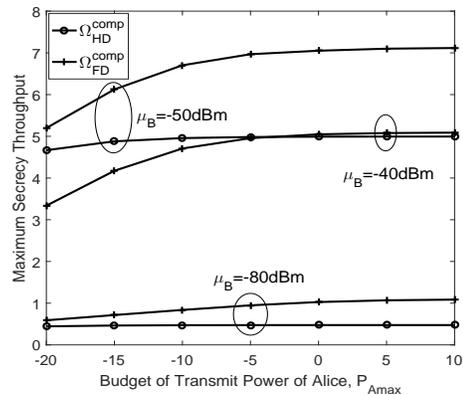}
\end{center}
\caption{Maximum secrecy throughput $\Omega_{FD}^{comp}$ and $\Omega_{HD}^{comp}$ vs. $P_{Amax}$ (dBm) for different $\mu_B's$ with $\lambda_e=10^{-5}, \epsilon=0.05, P_{Bmax}=10$dBm and $\rho=-70$dBm.}\label{Fig1HFc}
\end{figure}

Fig. \ref{Fig1HFc} plots the maximum secrecy throughputs $\Omega_{FD}^{comp}$ and $\Omega_{HD}^{comp}$ solved from \eqref{OFDcomp} and \eqref{OHDcomp}, respectively, versus $P_{Amax}$ for different values of $\mu_B$. $\Omega_{FD}^{comp}$ is larger than $\Omega_{HD}^{comp}$ when $\mu_B$ is small, since the jamming signal transmitted by Bob interferes with Eves. With increasing $\mu_B$, the superiority to Bob in the FD mode first increases and then decreases as the figure shows. $\Omega_{FD}^{comp}$ is smaller than $\Omega_{HD}^{comp}$ when $\mu_B$ is large enough, which confirms Bob stopping jamming when SI is dominant. For two curves of $\mu_B=-40dBm$, whether $\Omega_{FD}^{comp}$ is superior to $\Omega_{HD}^{comp}$ depends on $P_{Amax}$. We should note that the least $P_B$ is predefined as $-10dBm$ above, or otherwise the optimum $P_B$ is zero for sufficiently large $\mu_B$ or small $P_{Amax}$ and thus the FD mode degenerates into the HD mode.

\begin{figure}[!t]
\begin{center}
\centering
\includegraphics[height=5.3cm,width=7.1cm]{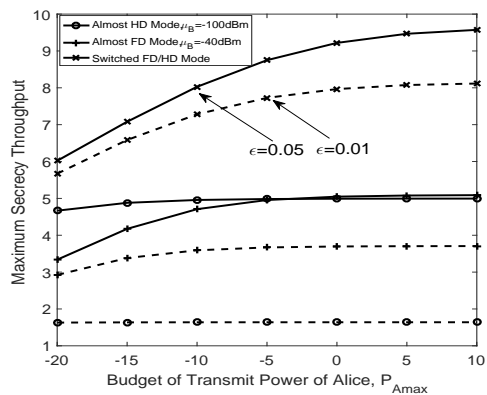}
\end{center}
\caption{Maximum secrecy throughput of three transmission schemes vs. $P_{Amax}$ (dBm) for different $\epsilon's$ with $\lambda_e=10^{-5}, P_{Bmax}=10$dBm and $\rho=-70$dBm.}\label{Fig1Switch}
\end{figure}

Fig. \ref{Fig1Switch} plots the maximum secrecy throughputs for the almost FD Bob, almost HD Bob, and switched FD/HD Bob all solved from our proposed scheme versus $P_{Amax}$ for different values of $\epsilon$. Obviously, the secrecy throughput with switched FD/HD Bob is always larger than the ones for FD and HD Bob. We should note that with $\mu_B=-40dBm$ and $\rho=-70dBm$, i.e., $\gamma_{BB} \leq 10^{3}$, Bob almost only operates in the FD mode considering $\gamma_{BB}$ distributed as $\exp \left(1\right)$. On the other hand, with $\mu_B=-100dBm$ and $\rho=-70dBm$, Bob almost operates in the HD mode only.

\begin{figure}[!t]
\begin{center}
\centering
\includegraphics[height=5.3cm,width=6.8cm]{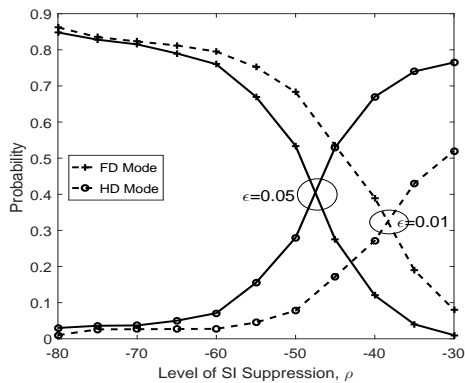}
\end{center}
\caption{Probability of Bob operating in the FD mode $\mathcal{P}_{FD}$ or the HD mode $\mathcal{P}_{HD}$ vs. $\rho$ (dBm) for different $\epsilon's$ with $\lambda_e=10^{-5}$, $P_{Amax}=10$dBm and $P_{Bmax}=30$dBm.}\label{Fig1Pro}
\end{figure}

Fig. \ref{Fig1Pro} illustrates the probability of Bob operating in the FD mode $\mathcal{P}_{FD}$ solved from \eqref{ProFD} and in the HD mode $\mathcal{P}_{HD}$ solved from \eqref{ProHD} of our proposed transmission scheme change with $\rho$ for different values of $\epsilon$. In accordance with the former analysis, $\mathcal{P}_{FD}$ decreases with a worse suppression level of SI, while $\mathcal{P}_{HD}$ is totally opposite. Moreover, $\mathcal{P}_{FD}$ decreases and $\mathcal{P}_{HD}$ increases w.r.t. $\epsilon$.

\begin{figure}
\centering
\subfigure[]{
\label{Fig1HybridPA} 
\includegraphics[height=5.5cm,width=7cm]{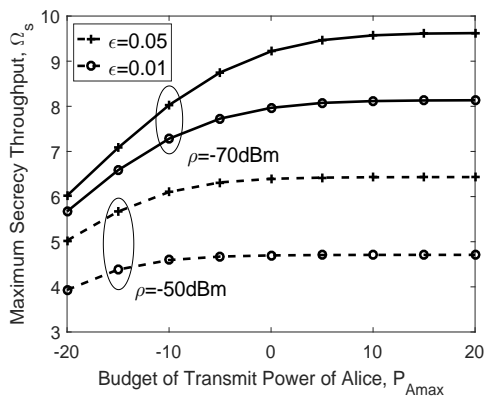}}
\hspace{0.1in}
\subfigure[]{
\label{Fig2HybridPB} 
\includegraphics[height=5.5cm,width=7cm]{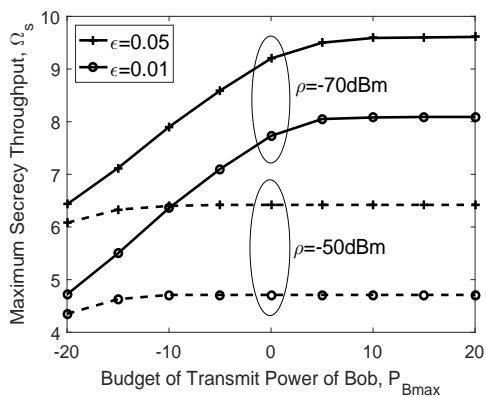}}
\caption{Maximum secrecy throughput $\Omega_s^{\star}$  vs. $P_{Amax}$ (dBm) with $P_{Bmax}=10$dBm (left) and $P_{Bmax}$ (dBm) with $P_{Amax}=10$dBm (right) for different values of $\epsilon$ and $\rho$ with $\lambda_e=10^{-5}$.}
\label{figHybrid} 
\end{figure}

Fig. \ref{figHybrid} interprets how the maximum secrecy throughput $\Omega_s^{\star}$ changes with $P_{Amax}$ and $P_{Bmax}$ for different values of $\epsilon$ and $\rho$. $\Omega_s^{\star}$ decreases with increasing $\rho$ since the secrecy performance is damaged by badly suppressed SI. As shown in Fig. \ref{Fig1HybridPA}, when $P_{Amax}$ grows, $\Omega_s^{\star}$ increases until it approaches a static value. Furthermore, the minimum $P_{Amax}$ at which $\Omega_s^{\star}$ stays unchanged (turning point) depends on the larger one between the two turning points' $P_{Amax}$ of the secrecy throughput for FD and HD Bob, respectively, since the ones of $\Omega_{FD}^\dag$ and $\Omega_{HD}^\ast$ are irrelevant to $\mu_B$ seen from Fig. \ref{Fig1HFc}. As shown in Fig. \ref{Fig2HybridPB}, $\Omega_s^{\star}$ increases until it approaches a static value with the growth of $P_{Bmax}$. Moreover, $P_{Bmax}$ of $\Omega_s^{\star}$'s turning point is practically the same as the one of $\Omega_{FD}^\dag$, since $P_{Bmax}$ is irrelevant to $\mu_B$ and $P_{Bmax}$ has no impact on the secrecy performance for HD Bob. Therefore, to enhance the secrecy throughput, we are inspired to rationally relax upper bound of SOP or to expand power budgets of Alice and Bob.

\section{Conclusion}\label{SEC Conclude}

In this paper, we have provided a low complexity adaptive D2D secure transmission scheme with a switched FD/HD mode jamming receiver, according to the ICSIs of the main link and residual SI link, against PPP randomly distributed eavesdroppers. Two groups of optimized transceiver parameters for the FD and HD receiver, respectively, and
the optimum adaptive mode switch threshold  to maximize the secrecy throughput under the SOP constraint, have been obtained.
The algorithm is consisted of the off-line and on-line parts, and  the computational complexity of the on-line operations is low. Simulation results show that the secrecy throughput increases with the growth of budgets of transmit power and jamming power respectively, until it monotonically reaches a plateau. This indicates an existing balance between the reliable connection probability and the SOP. Numerical comparisons on the secrecy throughput verify the superiority of the proposed switched FD/HD mode receiver over an FD or HD mode receiver. Moreover, the secrecy throughput would be enhanced with a larger suppression level of SI, sparser eavesdroppers or a less rigorous SOP constraint.

\appendices

\section{Proof of Lemma \ref{CDF of SINRe}}
\label{Appendix.A}
Considering the SINR $\varphi_E$ of \eqref{CE}, we have
\begin{align}
&\mathcal{F}_{\varphi_E}\left(x\right)\nonumber\\
&=\mathbb{P}\left\{
\max\frac{P_A\gamma_{Ak}d_{Ak}^{ - \alpha}}{\sigma_E^2+{P_B\gamma_{Bk}d_{Bk}^{ - \alpha}}}<x\right\}\nonumber\\
&=\mathbb{E}_\Phi\left[\mathop {\prod}\limits_{e_k\in\Phi}\mathbb{E}_{\gamma_{Bk}}\left[
\mathbb{P}\left\{\gamma_{Ak}<x\frac{\sigma_E^2+{P_B\gamma_{Bk}d_{Bk}^{-\alpha}}}{P_Ad_{Ak}^{-\alpha}} |\gamma_{Bk} \right\}\right] \right]\nonumber\\
&\overset{\left(d\right)}{=} \mathbb{E}_\Phi\left[\mathop{\prod}\limits_{e_k\in\Phi}\mathbb{E}_{\gamma_{Bk}}\left[
1-\exp\left(-x\frac{\sigma_E^2+{P_B\gamma_{Bk}d_{Bk}^{-\alpha}}}{P_Ad_{Ak}^{-\alpha}}\right) \right]\right]\nonumber\\
&\overset{\left(e\right)}{=} \mathbb{E}_\Phi\Bigg[\mathop{\prod}\limits_{e_k\in\Phi}\int\nolimits_{0}^{+\infty}
\exp\left(-\gamma_{Bk}\right) \nonumber\\
& \quad \quad \quad\quad\quad \times \left(
1-\exp\left(-x\frac{\sigma_E^2+{P_B\gamma_{Bk}d_{Bk}^{-\alpha}}}{P_Ad_{Ak}^{-\alpha}}\right) \right) d\gamma_{Bk}\Bigg]\nonumber\\
&\overset{\left(f\right)}{=} \exp\Bigg(-\lambda_e\int\nolimits_{0}^{2\pi} \int\nolimits_{0}^{+\infty}
\left(1+ x\frac{P_B}{P_A}\frac{d_{Ak}^\alpha}{d_{Bk}^\alpha}\right) ^ {-1}\nonumber\\
&\quad\quad\quad\quad\quad\quad\times \exp\left(-x\frac{\sigma_E^2}{P_A}d_{Ak}^\alpha\right)
d_{Ak}d{d_{Ak}}d\theta_k\Bigg), \label{CDF0}
\end{align}
where $\left(d\right)$ and $\left(e\right)$ follow from $\gamma_{Ak}\thicksim\exp\left(1\right)$ and $\gamma_{Bk}\thicksim\exp\left(1\right)$, respectively, and $\left(f\right)$ holds for the probability generating functional lemma (PGFL) over PPP \cite{SG1996}.

\section{Proof of Proposition \ref{p1FD}}
\label{Appendix.B}

Case 1: If $\mu_{A1}\left( R_C \right) \geq \mu_{A2}\left( R_C,R_S \right)$, we can obtain $\mathcal{G}\left( \mu_{A1}\left( R_C \right), \frac{\mu_B}{\rho}, R_C,R_S\right)\leq \tau$ according to $\mu_{A2}\left( R_C,R_S \right)$ in \eqref{mu2FD}. Combining with $\mu_{A1}\left( R_C \right)$ in \eqref{mu1FD}, we then obtain the following inequality
\begin{equation}
\left( \frac{\left( 2^{R_C-R_S}-1 \right)P_B }{P_{Amax}}+1\right)^{ - 1}\left( \frac{\left( 2^{R_C-R_S}-1 \right)\sigma_E^2}{P_{Amax}} \right)^{-\frac{2}{\alpha}}\leq \tau, \label{muO1FD}
\end{equation}
and the objective function in \eqref{FD_A} is transformed into
\begin{equation}
\tilde{\Omega}=R_S\exp\left( -\frac{\left(2^{R_C}-1\right) \left( \sigma_B^2+ P_B\mu_B \right)}{P_{Amax}d_{AB}^{-\alpha}} \right).\label{FD2A}
\end{equation}
Obviously, $\tilde{\Omega}$ in \eqref{FD2A} is a decreasing function of $R_C$. To have a small $R_C$, the left-hand side of \eqref{muO1FD} is set to its maximum value $\tau$, which refers to $\mathcal{G}\left( \mu_{A1}\left( R_C \right), \frac{\mu_B}{\rho}, R_C,R_S\right)= \tau$, i.e.,  $\mu_{A1}\left( R_C \right) = \mu_{A2}\left( R_C,R_S \right)$ from \eqref{mu2FD}.

Case 2: If $\mu_{A1}\left( R_C \right) \leq \mu_{A2}\left( R_C,R_S \right) $, we can obtain $\mathcal{G}\left( \mu_{A1}\left( R_C \right), \frac{\mu_B}{\rho}, R_C,R_S\right) \geq \tau$ from \eqref{mu2FD}. Combining with $\mu_{A1}\left( R_C \right)$ in \eqref{mu1FD}, we then obtain the inequality
\begin{equation}
\left( \frac{\left( 2^{R_C-R_S}-1 \right)P_B }{P_{Amax}}+1\right)^{ - 1}\left( \frac{\left( 2^{R_C-R_S}-1 \right)\sigma_E^2}{P_{Amax}} \right)^{-\frac{2}{\alpha}} \geq \tau. \label{muO2FD}
\end{equation}
The objective function \eqref{FD_A} combining with $\mu_{A2}\left( R_C,R_S \right)$ in \eqref{mu2FD} is expressed by
\begin{equation}
\tilde{\Omega}=R_S\exp\left( - \mathcal{G}_1 ^{-1} \left( \tau, \frac{\mu_B}{\rho}, R_C,R_S \right) \right).\label{FD3A}
\end{equation}

We thus develop the relationship between $\tilde{\Omega}$ and $R_C$. The partial derivative of $\tilde{\Omega}$ w.r.t. $R_C$ is 
\begin{align}
\frac{\partial{\tilde{\Omega}}}{\partial{R_C}} &= - R_S\exp\left( - \mu_{A2}\left(R_C, R_S \right)\right)\frac{\partial{\mu_{A2}\left(R_C, R_S \right)}}{\partial{R_C}}\nonumber\\
&= R_S\mu_{A2}\left(R_C, R_S \right)\exp\left( - \mu_{A2}\left(R_C, R_S \right)\right) \nonumber\\
&\quad\times \left(\frac{1}{2^{R_C-R_S}-1}- \frac{1}{2^{R_C}-1} \right)\ln 2.
\end{align}
Clearly, $\frac{\partial{\tilde{\Omega}}}{\partial{R_C}}>0$ and $\tilde{\Omega}$ increases w.r.t. $R_C$. As a result, $R_C$ is expected to be large enough. The left-hand side of \eqref{muO2FD} is set to its minimum value $\tau$, which means $\mathcal{G}\left( \mu_{A1}\left( R_C \right), \frac{\mu_B}{\rho}, R_C,R_S\right)= \tau$, i.e.,  $\mu_{A1}\left( R_C \right) = \mu_{A2}\left( R_C,R_S \right)$ from \eqref{mu2FD}.

Through discussions of the two cases, we obtain equations in \eqref{muAFDproposition}. The proof is completed.


\section{Proof of Theorem \ref{t1FD}}
\label{Appendix.C}
From \eqref{muOFD}, we have $\frac{d Z\left( Y \right)}{d Y} = -\frac{Z\left( Y \right)}{Y}$ using the derivative rule for implicit functions. The first-order derivative of $\tilde{\Omega}$ w.r.t. $Y$ in \eqref{FD41A} is then
\begin{equation}
\frac{\partial{ \tilde{\Omega}}}{\partial{Y}}=\exp\left( - U Y\right)
\left(\frac{1}{\left(1+Y\right)\ln 2} - U \log_2\left( \frac{1+Y}
 {1+YZ\left( Y \right)} \right) \right).
\end{equation}
Assume there is a variable $Y^\ast$ allowing $\frac{\partial{ \tilde{\Omega}}}{\partial{Y}}|_{ Y=Y^\ast}=0$. That is, $Y^\ast$ satisfies the equality given by
\begin{equation}
\frac{1}{\left(1+Y\right)\ln 2} = U \log_2\left( \frac{1+Y}
 {1+YZ\left( Y \right)} \right).\label{Appc1}
\end{equation}

The second-order derivative of $\tilde{\Omega}$ w.r.t. $Y$ at $Y^\ast$ is
\begin{equation}
\frac{\partial^2 \tilde{\Omega}}{\partial Y^2}|_{ Y=Y^\ast} =- \frac{U \left(1+Y^\ast\right)+1}{\left(1+Y^\ast\right)^2 \ln 2}
\exp\left( - U Y^\ast\right).
\end{equation}
Clearly, $\frac{\partial^2 \tilde{\Omega}}{\partial Y^2}|_{ Y=Y^\ast}<0$, and $\tilde{\Omega}$ is a quasi-concave function of $Y$. Plugging \eqref{Appc1} into \eqref{muOFD} finishes the proof.

\section{Proof of Corollary \ref{c1FD}}
\label{Appendix.D}
1) Relationships between $R_C^{\ast}$, $\tilde{\Omega}^\ast$ and $\epsilon$, $\lambda_e$:

Considering $\frac{d V\left( Y^\ast \right)}{d Y^\ast}=\frac{ \left(1+V\left( Y^\ast \right)\right) \left(1+ U\left( 1+Y^\ast \right) \right) }{U\left( 1+Y^\ast \right)^2} > 0 $, the left side of \eqref{t1FDequ} decreases with $R_C^\ast$ with the other variables unchanged. When $\epsilon$ decreases or $\lambda_e$ increases, the left side of \eqref{t1FDequ} needs reducing, i.e., enhancing $R_C^{\ast}$, to ensure the equality. We thus obtain the relationships between $R_C^{\ast}$ and $\epsilon$, $\lambda_e$.

From \eqref{FD5A}, we find that the maximum $\tilde{\Omega}^\ast$ for a given $P_B$ decreases with increasing $R_C^\ast$. Therefore, $\tilde{\Omega}^\ast$ increases with increasing $\epsilon$ or decreasing $\lambda_e$.

2) Relationship between $R_C^{\ast}$, $\tilde{\Omega}^\ast$ and $P_{Amax}$:

Since $\tau$ is independent of $P_{Amax}$, $P_{Amax}$ increases to ensure the equation \eqref{t1FDequ} if $V\left( Y^\ast \right)$ increases. Therefore, $R_C^{\ast}$ increases with increasing $P_{Amax}$.

According to $Y^\ast > V\left( Y^\ast \right)$ from \eqref{Dfd}, the first-order derivative of $\tilde{\Omega}^\ast$ w.r.t. $P_{Amax}$ satisfies
\begin{equation}
 \frac{d \tilde{\Omega}^\ast}{d P_{Amax}}
 = \frac{\left( Y^\ast - V\left( Y^\ast \right) \right)\exp\left( - U Y^\ast \right)}{\left(1+V\left( Y^\ast \right)\right) \left(1+Y^\ast\right) P_{Amax}\ln 2}> 0.
\end{equation}
We verify that $\tilde{\Omega}^\ast$ increases w.r.t. $P_{Amax}$.

\section{Proof of Corollary \ref{c2FD}}
\label{Appendix.E}

1)When $\epsilon \rightarrow 1$ or $\lambda_e \rightarrow 0$: Referring to 1) in Appendix \ref{Appendix.D}, the optimum $Y^\ast$ is close to zero, and thus the secrecy throughput $\tilde{\Omega}^\ast \rightarrow \frac{1}{U\ln 2} $ from \eqref{FD5A} decreases w.r.t. $P_B$.

2)When $\epsilon \rightarrow 0$ or $\lambda_e \rightarrow \infty$: We find $V\left(Y^\ast\right) P_B \gg 1$ from \eqref{t1FDequ}, where $V\left(Y^\ast\right)$ increases with $Y^\ast$ and $P_B$
according to \eqref{Dfd} and 1) in Appendix \ref{Appendix.D}. If $P_B \gg 1$, $V\left(Y^\ast\right) \rightarrow Y^\ast$ and the equality in \eqref{constraintFD} is transformed into $\varpi W\left(Y^\ast\right)=U V\left(Y^\ast\right)$. If $V\left(Y^\ast\right) \gg 1$, $Y^\ast \gg 0$ and we also have $\varpi W\left(Y^\ast\right)=U V\left(Y^\ast\right)$ from \eqref{constraintFD}. Therefore, in such a limiting case, we have
\begin{equation}
\frac{V\left(Y^\ast\right)}{P_{Amax}} = \frac{1}{\frac{\sigma_B^2}{\eta \mu_B}-P_B}.\label{c3fd}
\end{equation}
Substituting \eqref{c3fd} into \eqref{t1FDequ}, we obtain
\begin{equation}
 \left(\frac{P_B}{\frac{\sigma_B^2}{\eta \mu_B}-P_B}+1\right)^{ - 1}\left( \frac{\sigma_E^2}{\frac{\sigma_B^2}{\eta \mu_B}-P_B} \right)^{-\frac{2}{\alpha}}= \tau.
\end{equation}
$P_B^\ast$ is the unique root of above equality and decreases with $\epsilon$ while increasing with $\lambda_e$. Hence, the the first-order derivative of $\tilde{\Omega}^\dag$ w.r.t. $P_B^\dag$ is written as
\begin{align}
 \frac{d \tilde{\Omega}^\dag}{d P_B^\dag}
 &= - \left( - \frac{\varpi}{U^\dag} + \frac{ U^\dag V^{ 2}\left(Y^\dag \right)\left(1+Y^\dag \right) }{P_{Amax} \left(1+V\left(1+Y^\dag \right)\right) } + \frac{\varpi}{U^\dag} + \varpi Y^\dag \right)\nonumber\\ &\quad  \times \frac{\exp\left( - U^\dag Y^\dag \right)}{U^\dag \left(1+Y^\dag \right) \ln 2} \nonumber\\
 &< 0.
\end{align}
The limiting secrecy throughput $\Omega^\dag$ decreases with $P_B^\dag$ here and we complete the proof.

\end{document}